\documentclass[10pt]{article}
\usepackage[utf8]{inputenc}

\usepackage{amsmath}
\usepackage{amsfonts}
\usepackage{amssymb}
\usepackage{makeidx}
\usepackage{graphicx}
\usepackage{subcaption}
\usepackage{listings}
\usepackage{apacite}
\usepackage{natbib}
\usepackage{indentfirst}

\usepackage{siunitx}
\usepackage{multirow}
\usepackage{booktabs}

\usepackage{setspace}

\usepackage{authblk}
\usepackage[pdfborder={0 0 0}]{hyperref}
\usepackage[margin=1in]{geometry}

\title{OPTION PRICING IN MARKETS WITH INFORMED TRADERS}

\author[a]{YUAN HU}
\author[b]{ABOOTALEB SHIRVANI}
\author[c]{STOYAN STOYANOV}
\author[d]{YOUNG SHIN KIM}
\author[e]{FRANK J. FABOZZI}
\author[f]{SVETLOZAR T. RACHEV}

\affil[a]{\small Texas Tech University\\
\url{yuan.hu@ttu.edu}}
\affil[b]{\small Texas Tech University\\
\url{abootaleb.shirvani@ttu.edu}}
\affil[c]{\small Charles Schwab Corporation\\
\url{stoyan.stoyanov@gmail.com}}
\affil[d]{\small Stony Brook University\\
\url{aaron.kim@stonybrook.edu}}
\affil[e]{\small EDHEC Business School\\
\url{frank.fabozzi@edhec.edu}}
\affil[f]{\small Texas Tech University\\
\url{zari.rachev@ttu.edu}}

\begin{document}

\maketitle

\noindent {\textbf {Abstract}}\ \ \ \ The objective of this  paper is to introduce the theory of option pricing for markets with informed traders within the framework of dynamic asset pricing theory. We introduce new models for option pricing for informed traders in complete markets where we consider traders with information on the stock price direction and stock return mean. The Black-Scholes-Merton option pricing theory is extended  for markets with informed traders, where price processes are following continuous-diffusions. By doing so, the discontinuity puzzle in option pricing is resolved. Using market option data, we estimate the implied surface of the probability for a stock upturn, the implied mean stock return surface, and implied trader information intensity surface.

\section{Introduction}
The theory of option pricing (TOP), developed in the seminal works of Black \& Scholes (1973) and Merton (1973), provides the theory of finance with the fundamentals to understand, model and apply the processes for pricing contingent claims. Several works provide a comprehensive exposition of TOP such as Cochrane (2001), Duffie (2001), Skiadas (2009), Campbell (2000), \c{C}elik (2012), and Munk (2013). Although it is impossible to overlook TOP’s enormous influence on the theory of finance and its applications, there are some limitations of the original formulation of TOP due to several restrictive premises of the theory that are inconsistent with the findings of empirical studies on asset pricing processes. On the empirical side, it has been found that there is long-range dependence in asset price time series, volatility clustering of asset returns, skewness of the distribution of asset returns, heavy tails of the distribution of asset returns, and multivariate tail dependencies in the vector of asset returns\footnote{See Lo \& MacKinley (1988), Rachev \& Mittnik (2000), Schoutens (2003), Cont (2001), Cont \& Tankov (2004), and Rachev et al. (2011).}. Asset returns exhibiting these attributes are inconsistent with the assumptions of TOP. On the theoretical side, the following assumptions are questionable: (1) market participants have symmetric information\footnote{See Brunnermeier (2001) and Kelly \& Ljungqvist (2012).}, (2) prices are unpredictable\footnote{See Campbell \& Yogo (2005), Boucher (2006), Ang \& Bekaert (2007), and Caporin et al. (2013).}, (3) asset prices are not driven by fractional processes exhibiting  long-range dependence\footnote{See Lo (1991), Campbell et al. (1997), Andersson (1998), Diebold \& Inoue (2001), Nielsen (2010), Caporale \& Gil-Alana (2014), \iffalse. Bender, Sottinen \& Valkeila (2006), Bjork \& Hult (2005),\fi Cheridito (2003), Cont (2005), \iffalse Coviello \& Russo (2006), \fi Comte \& Renault (1998), and Rostek (2009).}, and (4) markets do not exhibit chaotic, or  irrational  behavior\footnote{See Hsieh (1990), Jovanovic \& Schinckus (2013), Rubinstein (2001), Shiller (2003), and Daniel \& Titman (1999).\iffalse, and Bloxham (2016), and Alexander (2013)\fi}. Studies have questioned these assumptions.

There is a vast literature on asset pricing with asymmetric information, most notably the models proposed by both Kyle (1985) and Back (1992)\footnote{See also Back \& Baruch (2004)\iffalse Back, Cao, \& Willard (2000)\fi, Back \& Pedersen (1998), \iffalse Baruch(2002),\fi Caldentey \& Stacchetti (2010), Cho (2003), and Collin-Dufresne \&  Fos (2015).}.  Both models assume a market with a continuous-time risky asset and asymmetric information. In the Kyle model there are three financial agents: the market maker, an insider trader (who knows a payoff which will be revealed at a pre-specified future time), and an uninformed (noisy) trader. The market maker has to define a pricing rule in such a way that an equilibrium exists between the traders. Back (1993) extended the model to continuous time. A second line of research focuses on the study of markets with asymmetric information based on an enlargement of the filtration and the change of the probability measure, the study by Aase et al. (2010) being one example.

Our paper is close in spirit to Horst \& Naujokat (2011). In our paper, traders operate in an imperfect market.  The buyer and seller have different market information when making their option trades, but we deal with perfectly liquid markets. In Horst \& Naujokat (2011), the trades are executed in an illiquid market and option traders “manipulate” the option portfolio value by impacting the slippage in trading (hedging) the underlying. The common feature in both papers is that the option sellers use their hedged portfolio either as being more informed than the market (as in our paper), or “to increase their portfolio value by using their impact on the dynamics of the underlying” in the paper by Horst \& Naujokat (2011). Our approach is based on dynamic asset pricing theory, while Horst and Naujpkat employ an equilibrium pricing approach.

We derive option pricing formulas when some group of traders are in the possession of additional information about future asset prices. The information available to traders is multifaceted and any general definition will be restrictive in view of the traders’ particular trading activities. If traders have information on the stock price direction, we find that the fair option price follows the Black-Scholes-Merton formula with an additional term which can be interpreted as a continuous dividend stream. If traders have better information on the stock mean return, then the fair option price differs from the Black-Scholes-Merton formula only if continuous-time trading is not allowed.

The paper is organized as follows. In Section 2 we analyze the discontinuity puzzle in option pricing. Applying the option pricing model developed by Kim et al. (2019), we outline the resolution of the puzzle, and motivate our approach to the theory of option pricing for informed traders. We then estimate the implied surfaces of the probability for upturn and implied mean return on market data. In Section 3, we derive option pricing when the hedger has information about the stock's price direction. We estimate the implied surface of a trader who possesses information on the probability for a stock’s upturn. In Section 4, we extend the results in Section 3, deriving an option pricing model when the hedger has information about the stock mean return. In Section 5, the approach in Section 3 is generalized to cover markets with stock prices driven by continuous-diffusion processes.  Section 6 concludes the paper.

\section{The Discontinuous Option Pricing Model and Market with Informed Traders}
Our interest on the topic of option pricing in the presence of informed traders started with an attempt to remove an unnatural discontinuity of the derivative price valuation\footnote{See Kim et al. (2016, 2019), where the problem of discontinuity in option pricing was first discussed.}.

\subsection{The discontinuity puzzle in option pricing}
Consider the Black-Scholes-Merton market $(\mathcal{S},\mathcal{B},\mathcal{C})$ of risky asset (stock) $\mathcal{S}$, riskless asset\footnote{We also refer to $\mathcal{B}$ as a riskless bank account, or equivalently as a riskless bond.} $\mathcal{B}$, and European Contingent Claim (ECC) $\mathcal{C}$ . The stock price dynamic follows a geometric Brownian motion (GBM)
\begin{equation}
	S_t = S_t^{(\mu,\sigma)} = S_0e^{(\mu-\frac{1}{2}\sigma^2)t+\sigma B_t},t\geq 0,S_0>0,\mu >0,\sigma >0.
	\label{eq_price1}
\end{equation}
The GBM is on the natural world $\mathbb{P}$, determined by a stochastic basis $(\Omega,\mathbb{F} = \{\mathcal{F}_t\}_{t\geq 0},\mathbb{P})$ with filtration $\mathbb{F}$, generated by the Brownian motion (BM) $B_t,t \geq 0$. The bond price is given by
\begin{equation}
	\beta_t = \beta_0 e^{rt},t\geq 0,\beta_0 > 0,r \in (0,\mu).
	\label{eq_beta2}
\end{equation}
The ECC with underlying asset $\mathcal{S}$ has terminal (expiration) time $T>0$, and terminal payoff $g(S_T^{(\mu,\sigma)})$. Consider, as an example, a call option with strike price $K>0$. Then $g(S_T^{(\mu,\sigma)})= \max (S_T^{(\mu,\sigma)}-K,0)$. In the the Black-Scholes-Merton pricing framework, the call option price $C_t^{(K)} = C(S_t^{(\mu,\sigma)},K,T-t,r,\sigma)=C(S_t^{(r,\sigma)},K,T-t,r,\sigma)$ is independent of  the (instantaneous) stock mean return $\mu$. If $\mu \uparrow \infty$, the hedger (the trader taking the short position  in the option contract),  who \textit{can trade continuously in time with no transaction costs involved}\footnote{There is a vast literature on option price with transaction costs. Davis et al. (1993, 471) summarize the problems associated with continuous-time trading with no transaction costs.}, will be indifferent to the large values of $\mu$. Indeed, continuous-time trading with no transaction costs is a pure fiction in any real trading. Obviously, if $\mu = \infty$, $C(S_t^{(\infty,\sigma)},K,T-t,r,\sigma)=\infty$. However, by the Black-Scholes-Merton Theorem\footnote{See Black \& Scholes (1973) and Merton (1973).}, for every fixed $t\in [0,T)$, $ \sup_{\mu \in R} C(S_t^{(\mu,\sigma)},K,T-t,r,\sigma)=C(S_t^{(r,\sigma)},K,T-t,r,\sigma) < \infty$, and thus, we observe an unnatural discontinuity of the price of the option when $\mu \uparrow \infty$. Similarly, if $\mu = - \infty$, $C(S_t^{(- \infty,\sigma)},K,T-t,r,\sigma)= 0$. However according to the Black-Scholes-Merton framework, for every fixed $t\in [0,T),\; \inf_{\mu \in R} C(S_t^{(\mu,\sigma)},K,T-t,r,\sigma)=C(S_t^{(r,\sigma)},K,T-t,r,\sigma) > 0$, and thus, we again observe an unnatural discontinuity of the price of the derivative when  $\mu \downarrow - \infty$.

Therefore, we conclude that no real option trader will disregard the information about the mean stock return value $\mu$ in trading the option $\mathcal{C}$. That information should be embedded in option price $C_t^{(K)} = C(S_t^{(\mu,\sigma)},K,T-t,r,\sigma)$.

\subsection{The discontinuity in option pricing in binomial models}
Consider next the seminal Cox-Ross-Rubinstein (CRR) binomial option pricing model\footnote{See Cox, Ross, \& Rubinstein (1979) and Chapters 12 and 20 in Hull (2012) with $\mu = r$.}. The price process given  by (\ref{eq_price1}) generates a triangular series of $\Delta t$ log-returns, $r_{k \Delta t} = \log S_{k \Delta t}^{(\mu,\sigma)} - \log S_{(k-1) \Delta t}^{(\mu,\sigma)}, k \in \mathbb{N}_n = \{1,...,n\}, n \Delta t = T$. The returns are  independent and identically distributed  (\textit{i.i.d.}) Gaussian random variables with mean $\mu_{k \Delta t} = (\mu - \frac{1}{2}\sigma^2)\Delta t$, and variance $\sigma^2_{k \Delta t}=\sigma^2 \Delta t$, and denoted as $r_{k \Delta t} \overset{\mathrm{d}}{=} \mathcal{N}(\mu_{k \Delta t},\sigma^2_{k \Delta t})$.\footnote{Here and in what follows, $\overset{\mathrm{d}}{=}$ stands for “equal in distribution”, or “equal in probability law”.} Applying the CRR-model and the Donsker-Prokhorov Invariance Principle  (DPIP)\footnote{See Donsker (1951), Prokhorov (1956), Section 14 in Billingsley (1999), Chapter IX in Gikhman \& Skorokhod (1969), Section 5.3.3 in Skorokhod (2005), and Davydov \& Rotar (2008).}, we approximate $r_{k \Delta t}$ by
\begin{equation}
	r_{k \Delta t;n} = U_{\Delta t}^{(CRR)} \zeta_{k,n}^{(CRR)}+D_{\Delta t}^{(CRR)} (1-\zeta_{k,n}^{(CRR)}),k \in \mathbb{N}_n,
	\label{eq_return3}
\end{equation}
where
\begin{equation}
	U_{\Delta t}^{(CRR)} = \sigma \sqrt{\Delta t}, D_{\Delta t}^{(CRR)} = - \sigma \sqrt{\Delta t}
	\label{eq_up4}
\end{equation}
for every fixed $n \in \mathbb{N} = \{1,2,...\}$, and $\{ \zeta_{k,n}^{(CRR)}, k \in \mathbb{N}_n\}$ are \textit{i.i.d.} Bernoulli random variables, $\zeta_{k,n}^{(CRR)} \overset{\mathrm{d}}{=} Ber(p_{\Delta t}^{CRR}), \mathbb{P}(\zeta_{k,n}^{(CRR)} = 1) = 1-\mathbb{P}(\zeta_{k,n}^{(CRR)} = 0) = p_{\Delta t}^{(CRR)} \in (0,1)$ with success probability\footnote{Under (\ref{eq_prob_5}), the risk -neutral probability in the CRR-binomial pricing model is $q_{\Delta t}^{(CRR)} = \frac{e^{r \Delta t}-e^{-\sigma \sqrt{\Delta t}}}{e^{\sigma \sqrt{\Delta t}}-e^{-\sigma \sqrt{\Delta t}}} = \frac{1}{2}+\frac{r-\frac{\sigma^2}{2}}{2\sigma}\sqrt{\Delta t}$ with $o(\Delta t) = 0$, see Kim et al. (2016, 2019).}
\begin{equation}
	p_{\Delta t}^{(CRR)} = \frac{e^{\mu \Delta t}-e^{-\sigma \sqrt{\Delta t}}}{e^{\sigma \sqrt{\Delta t}}-e^{-\sigma \sqrt{\Delta t}}}.
	\label{eq_prob_5}
\end{equation}

A widely used alternative to the CRR binomial pricing tree is the Jarrow-Rudd (JR) binomial model\footnote{See Jarrow \& Rudd (1983, p. 179-190) and Section 20.4 in Hull (2012) with $\mu = r$.}, where in (\ref{eq_return3}), (\ref{eq_up4}) and (\ref{eq_prob_5}), the triplet $(U_{\Delta t}^{(CRR)},D_{\Delta t}^{(CRR)},p_{\Delta t}^{(CRR)})$ is replaced by triplet
$(U_{\Delta t}^{(JR)},D_{\Delta t}^{(JR)},p_{\Delta t}^{(JR)})$,
\begin{align}
	U_{\Delta t}^{(JR)}= (\mu - \frac{1}{2}\sigma^2)\Delta t + \sigma \sqrt{\Delta t},\; D_{\Delta t}^{(JR)} = (\mu - \frac{1}{2}\sigma^2)\Delta t - \sigma \sqrt{\Delta t},\; p_{\Delta t}^{(JR)} &= \frac{1}{2}
	\label{eq_tree_up6}
\end{align}
determining the JR-binomial tree\footnote{Under (\ref{eq_tree_up6}), the risk-neutral probability in the JR-binomial pricing model is $q_{\Delta t}^{(JR)} = \frac{1}{2}-\frac{1}{2}\theta \sqrt{\Delta t}$ with $o(\Delta t) = 0$, where $\theta = \frac{\mu - r}{\sigma}$ is the market price of risk. If the risk-neutral probability $q_{\Delta t}^{(JR)} = \frac{1}{2}$, is as in the original JR-model, then the corresponding natural-world-probability $p_{\Delta t}^{(JR)}$ can be either $p_{\Delta t}^{(JR)} = \frac{1}{2}-\frac{1}{2}\theta \sqrt{\Delta t}$, or $p_{\Delta t}^{(JR)} = \frac{1}{2}+\frac{1}{2}\theta \sqrt{\Delta t}$, see Kim et at (2016, 2019).}. The CRR- and JR-binomial pricing tree constructions have many advantages, unfortunately, they have one common disadvantage – an option-price-discontinuity. To illustrate that,  consider a one-period binomial model, where
\begin{itemize}
	\item[1.]$S_0 >0$ is the known current (at $t=0$) one-share stock price, 
	\item[2.]$f_0$ is the unknown current option price,
	\item[3.]$S_T = \begin{cases} 
      S_0 u & w.p. \; p_0\\
      S_0 d & w.p. \; 1-p_0 
   \end{cases},$ where $p_0 \in (0,1)$, is the stock price at the option's maturity T for some $u>0$ and $d>0$, satisfying the no-arbitrage condition $u>e^{rT}>d$,
	\item[4.]the option payoff at maturity is
\begin{equation}
	f_T = \begin{cases} 
      f_T^{(u)} = g(S_0 u) & w.p. \; p_0 \\
      f_T^{(d)} = g(S_0 d) & w.p. \; 1-p_0 
   \end{cases}
   \label{eq_payoff7}
\end{equation}
for some option payoff function $g(x),x>0$.
\end{itemize}
For any $p_0 \in (0,1)$, the value of the option at $t=0$ is given by $f_0(p_0) = f_0(\frac{1}{2}) = e^{-rT}[q_0f_T^{(u)}+(1-q_0)f_T^{(d)}]$ with $q_0 = \frac{e^{rT}-d}{u-d}$,\footnote{See, for example, Hull (2012, p.256).} regardless of how close $p_0$ is to $0$ or $1$. However, for $p_0 = 0$ and $p_0 = 1$, the option values are, respectively, $f_0(0) = e^{-rT} f_T^{(d)}$ and $f_0(1) = e^{-rT} f_T^{(u)}$. The discontinuity gaps at $p_0 = 0$ and $p_0 = 1$, are respectively,
\begin{equation}
	\begin{cases} 
      f_0(0)-\lim_{p \downarrow 0} f_0(p_0) =  e^{-rT}q_0(f_T^{(d)}-f_T^{(u)})\neq 0,\\
      f_0(1)-\lim_{p \uparrow 1} f_0(p_0) =  e^{-rT}(1-q_0)(f_T^{(u)}-f_T^{(d)})\neq 0.
   \end{cases}
   \label{eq_tree8}
\end{equation}

In contrast to the discontinuity gaps
\begin{equation}
	\begin{cases} 
		C(S_t^{(-\infty,\sigma)},K,T-t,r,\sigma) - \inf_{\mu \in R} C(S_t^{(\mu,\sigma)},K,T-t,r,\sigma) <0\\
		C(S_t^{(-\infty,\sigma)},K,T-t,r,\sigma) - \sup_{\mu \in R} C(S_t^{(\mu,\sigma)},K,T-t,r,\sigma) = \infty
	\end{cases}
	\label{eq_option9}
\end{equation}
reported in Section 2.1, where (\ref{eq_option9}) could be explained by the (presumed) hedger's ability to trade continuously in time with no transaction cost, in (\ref{eq_tree8}), the discontinuity gaps are present in one-period binomial pricing, and that makes the issue of option-price-discontinuity even more disturbing.

The main reason for the discontinuity phenomenon (\textit{the discontinuity puzzle in option pricing}) in (\ref{eq_tree8}) and (\ref{eq_option9}) is that a trader $\aleph_0$,\footnote{We will designate this trader as a \textit{noise trader} $\aleph_0$.} taking a short position in the option contract, is applying the CRR model disregarding any information about the  mean return $\mu$ and probability for stock upturn $p_0 \in (0,1)$. However, at $p_0 =1$, the trader becomes fully aware that the stock price will be up; that is, this trader becomes a \textit{trader with complete information}\footnote{We will designate this trader, as a \textit{fully informed trader} $\aleph_{\infty}$.} about the stock price direction. This jump from a noisy trader $\aleph_0$ to a fully informed trader $\aleph_{\infty}$ seems unnatural.

To resolve this issue, in this paper we will assume that the trader, called $\aleph$, knows, at time $t = k \Delta t, k = 0,..., n-1, n\Delta t =T$, with certain probability $p_{\Delta t}^{\aleph} \in (0,1)$, the correct (true) stock direction in $[k \Delta t, (k+1) \Delta t]$, or has information about the mean $\mu$. We then have that $\aleph$ is an \textit{informed trader} if $p_{\Delta t}^{\aleph} > \frac{1}{2}$, a \textit{misinformed trader} if $p_{\Delta t}^{\aleph} < \frac{1}{2}$, and a \textit{noisy trader} if $p_{\Delta t}^{\aleph} = \frac{1}{2}$. To illustrate our approach in this paper, assume that $\aleph$ is an informed trader who knows (at $t=0$) with probability $p^{\aleph} \in (\frac{1}{2},1)$ of the stock price direction at $t = T$. The stock price at $T$ is given by
\begin{align*}
S_T = \begin{cases} 
      S_0 u & w.p. \; p_0 \\
      S_0 d & w.p. \; 1-p_0
\end{cases},\; p_0 \in (0,1).
\end{align*} 
$\aleph$ chooses $u$ and $d$, so that the stock return $R_T = \frac{S_T}{S_0}-1$ has mean $\mathbb{E}(R_T) = \mu_T T$,\footnote{We assume $\mu_T>r_T>0$, where $r_T>0$ is the riskless rate in $[0,T]$. The no-arbitrage condition requires $u>1+r_T T>d$.} and variance $Var(R_T) = \sigma_T^2 T,\sigma_T>0$. The two moment conditions lead to $u = 1+\mu_T T+\sigma_T \sqrt{\frac{1-p_0}{p_0}T}$ and $d = 1+\mu_T T-\sigma_T \sqrt{\frac{p_0}{1-p_0}T}$. Next, consider the option on the stock with price $f_0$ at $t=0$, and payoff at maturity $t = T$ given by (\ref{eq_payoff7}). The  self-financing portfolio comprised of a stock and bond\footnote{Without loss of generality, we assume $\beta_0=1$ in (\ref{eq_beta2}).}, replicating the option value is $P_0 = a_0S_0+b_0 = f_0$. Then, $f_T = a_0 S_T+b_0(1+r_T T)$. By the risk-neutrality, $a_0S_0u+b_0(1+r_T T)-f_T^{(u)} = a_0 S_0 d+b_0(1+r_T T)-f_T^{(d)} = 0$. This leads to $a_0 = \frac{f_T^{(u)}-f_T^{(d)}}{S_0(u-d)}$ and $b_0 = \frac{1}{1+r_T T} \frac{f_T^{(d)}u-f_T^{(u)}d}{u-d}$. The option price at $t=0$ is given by $f_0 = \frac{1}{1+r_T T}(q_0 f_T^{(u)}+(1-q_0)f_T^{(d)})$, where $q_0 = p_0-\theta_T \sqrt{p_0(1-p_0)T}$ is the risk-neutral probability, and $\theta_T = \frac{\mu_T-r_T}{\sigma_T}$ is the market price of risk. 

Suppose $\aleph$ takes a short position in the option contract with terminal payoff (\ref{eq_payoff7}). If, at $t=0$, $\aleph$ believes that the stock will move “upward”, he enters an $N^{\aleph}$ long forward contract. If, at $t=0$, $\aleph$ believes that the stock will move “downward”, he enters an $N^{\aleph}$ short forward contract. The probability of  $\aleph$ being correct in his guess on the stock price direction is $p^{\aleph} \in (\frac{1}{2},1)$. As it costs nothing to enter the  forward contract, $\aleph$ replicates his short position in the option with the price process: $S_0^{\aleph} = S_0$ and 
\begin{align*}
	S_T^{\aleph} = \begin{cases} 
      S_0 u + N^{\aleph}(S_0 u - S_0(1+r_T T)) & w.p. \; p_0 p^{\aleph}, \\
      S_0 d + N^{\aleph}(S_0 (1+r_T T) - S_0 d) & w.p. \; (1-p_0) p^{\aleph},  \\
      S_0 u + N^{\aleph}(S_0 (1+r_T T) - S_0 u) & w.p. \; p_0 (1-p^{\aleph}),  \\
      S_0 d + N^{\aleph}(S_0 d - S_0(1+r_T T)) & w.p. \; (1-p_0) (1-p^{\aleph}).  
	\end{cases}
\end{align*}
Then the mean and the variance of the stock return $R^{\aleph}_T = \frac{S^{\aleph}_T}{S^{\aleph}_0}-1$ is given by
\begin{align*}
	\mathbb{E}(R^{\aleph}_T) &= \mu_T T +\sigma_T \sqrt{T} N^{\aleph} (2p^{\aleph}-1)(\theta_T \sqrt{T}(2p_0-1)+2 \sqrt{p_0(1-p_0)}),\\
	Var(R^{\aleph}_T) &= \sigma_T^2 T[1+N^{\aleph^2}+N^{\aleph^2}\theta_T^2 T + 2N^{\aleph}(2p^{\aleph}-1)(2\theta_T \sqrt{p_0(1-p_0)T}+1-2p_0)] \\
	&- \sigma_T^2 T [N^{\aleph^2}(2p^{\aleph}-1)^2(\theta_T\sqrt{T}(2p_0-1)+2 \sqrt{p_0(1-p_0)})^2].
\end{align*}

To simplify  the exposition in this example, we set $T = \Delta t$, with $o(\Delta t) = 0$. Then, $\mu_{\Delta t} = \mu$, $\sigma_{\Delta t} = \sigma$, and $r_{\Delta t} = r$. Assume that $p^{\aleph} = p^{\aleph}_{\Delta t} = \frac{1}{2}(1+\psi^{\aleph} \sqrt{\Delta t})$ for some $\psi^{\aleph}>0$.\footnote{$\psi^{\aleph}$ is  $\aleph$'s \textit{stock price direction information intensity}.} Then $\mathbb{E}(R_{\Delta t}^\aleph) = (\mu+2N^\aleph \sigma \sqrt{p_0(1-p_0)}\psi^{\aleph})\Delta t$ and $Var(R_{\Delta t}^{\aleph}) = \sigma^2(1+N^{\aleph^2})\Delta t$. $\aleph$ determines the optimal $N^{\aleph} = N^{(\aleph;opt)}$ as the one that maximizes the instantaneous market price of risk $\Theta(R_{\Delta t}^{\aleph}) = \frac{\mathbb{E}(R_{\Delta t}^\aleph)-r\Delta t}{\sqrt{Var(R_{\Delta t}^\aleph)\Delta t}} = \frac{\theta+2 N^\aleph \sqrt{p_0(1-p_0)}\psi^\aleph}{\sqrt{1+N^{\aleph^2}}},\theta = \frac{\mu-r}{\sigma}>0$. Choosing $N^{\aleph} = N^{(\aleph;opt)} = \frac{2 \psi^{\aleph}\sqrt{p_0(1-p_0)}}{\theta_{\Delta t}}$ leads to $\Theta(R_{\Delta t}^{\aleph}) = \Theta^{(opt)}(R_{\Delta t}^{\aleph}) = \sqrt{\theta^2+4\psi^{\aleph^2}p_0(1-p_0)}$. Furthermore, the optimal mean and variance of the return $R_{\Delta t}^{\aleph}$ are $\mathbb{E}(R_{\Delta t}^\aleph) = \mu ^\aleph \Delta t$ and $Var(R_{\Delta t}^\aleph) = \sigma^{\aleph^2} \Delta t$, where $\mu^\aleph = \mu+4 \sigma p_0(1-p_0) \frac{\psi^{\aleph^2}}{\theta}$ and $\sigma^\aleph = \sigma \sqrt{1+4p_0(1-p_0)\frac{\psi^{\aleph^2}}{\theta^2}}$. Next, $\aleph$ hedges the stock price movements, upward and downward, using stock price process\footnote{$\aleph$ does not hedge the risk of his bet on the stock price direction being wrong. He hedges only the risk of stock’s upward or downward movements.} 
\begin{align*}
S_{\Delta t}^{(\aleph;opt)} = 
\begin{cases} 
      S_0 u_{\Delta t}^\aleph & w.p. \; p_0\\
      S_0 d_{\Delta t}^\aleph & w.p. \; 1-p_0 
\end{cases},
\end{align*}
where $u_{\Delta t}^\aleph = 1+\mu^\aleph \Delta t+\sigma^\aleph \sqrt{\frac{1-p_0}{p_0}\Delta t}$ and $d_{\Delta t}^\aleph = 1+\mu^\aleph \Delta t-\sigma^\aleph \sqrt{\frac{p_0}{1-p_0}\Delta t}$. For $\aleph$, the option price is now $f_0^\aleph = \frac{1}{1+r \Delta t}(q_0^\aleph f_{\Delta t}^{(u)}+(1-q_0^\aleph)f_{\Delta t}^{(d)})$, where $q_0^\aleph = p_0-\theta^{\aleph}\sqrt{p_0(1-p_0)\Delta t}$, and $\theta^{\aleph} = \frac{\mu^\aleph-r}{\sigma^\aleph} = \sqrt{\theta^2+4p_0(1-p_0)\psi^{\aleph^2}}$. This results in an option price when the underlying stock is paying dividend $D_y^\aleph>0$. $\aleph$ receives the dividend yield  $D_y^\aleph$ making use of his information about the stock's price movement. The yield $D_y^\aleph$ is determined by $\theta^{\aleph} = \frac{\mu^\aleph-r}{\sigma^\aleph}= \frac{\mu+D_y^\aleph-r}{\sigma}$, and is equal to $D_y^\aleph = \sigma(\sqrt{\theta^2+4p_0(1-p_0)\psi^{\aleph^2}}-\theta)$. If $\aleph$ is a misinformed trader, he does just the opposite of an informed trader, and what will be a profit  for the informed trader will be a loss for the misinformed trader. Thus, in general, if $p^\aleph = p^\aleph_{\Delta t} = \frac{1}{2}(1+\psi^\aleph \sqrt{\Delta t})$ for some $\psi^\aleph \in R$, the yield $D_y^\aleph \in R$, is given by $D_y^\aleph = \textup{sign}(\psi^\aleph)\sigma(\sqrt{\theta^2+4p_0(1-p_0)\psi^{\aleph^2}}-\theta)$, where 
\begin{align*}
\textup{sign}(\psi^\aleph) = 
\begin{cases} 
      1, & if \; \psi^\aleph >0 \\
      0, & if \; \psi^\aleph =0 \\
      -1, & if \; \psi^\aleph <0 \\
\end{cases}.
\end{align*}
We will elaborate on this approach to option pricing for informed traders in Sections 3, 4 and 5.

\subsection{KSRF binomial option pricing}
\par{\fontsize{10}{13}In this section we provide a summary of the Kim-Stoyanov-Rachev-Fabozzi (KSRF) binomial option pricing (Kim et al., 2016 and 2019) which will be used in this paper as a basic model for discrete asset pricing.}

Consider again, a market of three assets: risky asset (stock) $\mathcal{S}$, riskless asset (riskless bank account, riskless bond) $\mathcal{B}$, and a derivative (option) $\mathcal{C}$. In continuous time, the stock price dynamics $S_t = S_t^{(\mu,\sigma)},t \in [0,T]$ is given by (\ref{eq_price1}). The bond price is given by (\ref{eq_beta2}), and the option contract $\mathcal{C}$ has continuous price process $f_t = f(S_t,t), t\in [0,T)$, and terminal payoff, $f_T = g(S_T)$, where the real-valued  function $f(x,t),x>0,t\in[0,T)$ is sufficient smooth. The log-returns $r_{k \Delta t} = \log S_{k \Delta t}^{(\mu,\sigma)} - \log S_{(k-1) \Delta t}^{(\mu,\sigma)}, k \in \mathbb{N}_n = \{1,...,n\}, n \Delta t = T$ are \textit{i.i.d.} Gaussian random variables $r_{k \Delta t} \overset{\mathrm{d}}{=} \mathcal{N}((\mu-\frac{\sigma^2}{2})\Delta t,\sigma^2 \Delta t)$. Following CRR binomial pricing model's construction, KSRF introduce their binomial pricing tree. Consider the discrete filtration $\mathbb{F}^{(n)} = \{\mathcal{F}_{k;n} = \sigma(\zeta^{(p_{\Delta t})}_{1,n},...,\zeta^{(p_{\Delta t})}_{k,n}), k \in \mathbb{N}_n, \mathcal{F}_{0;n}=\{\varnothing,\Omega \} \}$, where $\{\zeta_{k,n}^{(p_{\Delta t})}, k \in \mathbb{N}_n\}$ are \textit{i.i.d.} Bernoulli random variables with $\mathbb{P}(\zeta^{(p_{\Delta t})}_{k,n} = 1) = 1-\mathbb{P}(\zeta^{(p_{\Delta t})}_{k,n} = 0) = p_{\Delta t} \in (0,1)$. The KSRF binomial pricing tree is defined as follows $S_{0,n}^{(p_{\Delta t})} = S_0$, and for $k = 1,...,n-1$, conditionally on $\mathcal{F}_{k;n}$,
\begin{equation}
	S_{k+1,n}^{(p_{\Delta t})} = 
	\begin{cases} 
      S_{k+1,n}^{(p_{\Delta t},u)} = S_{k,n}^{(p_{\Delta t})}e^{U_{\Delta t}},& \textrm{if} \; \zeta_{k+1,n}^{(p_{\Delta t})} = 1 \\
      S_{k+1,n}^{(p_{\Delta t},d)} = S_{k,n}^{(p_{\Delta t})}e^{D_{\Delta t}},& \textrm{if} \; \zeta_{k+1,n}^{(p_{\Delta t})} = 0
	\end{cases}
	= S_{k,n}^{(p_{\Delta t})} 
	\begin{cases} 
      e^{U_{\Delta t}}, & w.p.\; p_{\Delta t}\\
      e^{D_{\Delta t}}, & w.p.\; 1-p_{\Delta t}
	\end{cases},
	\label{eq_option10}
\end{equation}
where
\begin{equation}
	\begin{cases} 
      U_{\Delta t} = (\mu-\frac{\sigma^2}{2}\frac{1-p_{\Delta t}}{p_{\Delta t}})\Delta t+\sigma \sqrt{\frac{1-p_{\Delta t}}{p_{\Delta t}}}\sqrt{\Delta t}\\
      D_{\Delta t} = (\mu-\frac{\sigma^2}{2}\frac{p_{\Delta t}}{1-p_{\Delta t}})\Delta t-\sigma \sqrt{\frac{p_{\Delta t}}{1-p_{\Delta t}}}\sqrt{\Delta t}
	\end{cases}.
	\label{eq_option11}
\end{equation}
With $o(\Delta t) = 0$, the binomial tree (\ref{eq_option10}), has the equivalent form
\begin{equation}
	S_{k+1,n}^{(p_{\Delta t})} = \begin{cases} 
	S_{k+1,n}^{(p_{\Delta t},u)} = S_{k,n}^{(p_{\Delta t})}(1+\mu\Delta t+\sigma \sqrt{\frac{1-p_{\Delta t}}{p_{\Delta t}}}\sqrt{\Delta t}),& \textrm{if} \; \zeta^{(p_{\Delta t})}_{k+1,n} = 1\\
	S_{k+1,n}^{(p_{\Delta t},d)} = S_{k,n}^{(p_{\Delta t})}(1+\mu\Delta t-\sigma \sqrt{\frac{p_{\Delta t}}{1-p_{\Delta t}}}\sqrt{\Delta t}),& \textrm{if} \; \zeta^{(p_{\Delta t})}_{k+1,n} = 0
	\end{cases}.
	\label{eq_tree12}
\end{equation}
If $p_{\Delta t} = p_{\Delta t}^{(CRR)}$, and $o(\Delta t) = 0$, the KSRF pricing tree (\ref{eq_option10}) and (\ref{eq_option11}) becomes the CRR-pricing tree (\ref{eq_return3}) and (\ref{eq_up4}). If $p_{\Delta t} = \frac{1}{2}$, the KSRF pricing tree becomes the JR-pricing tree. By the DPIP, the $\mathcal{D}[0,T]$-process, $\mathbb{S}^{(n)} = \{S_t^{(n)} = S_t^{(n;\mu,\sigma)} = S_{k,n}^{(p_{\Delta t})}, t \in [k\Delta t, (k+1)\Delta t),k=0,1,...,n-1, S_T^{(n)} = S_{n,n}^{(p_{\Delta t})}\}$ converges weakly in $\mathcal{D}[0,T]$ to $\mathcal{S} = \{S_t = S_t^{(\mu,\sigma)}, t\in [0,T]\}$. The discrete dynamics of $f_t,t \in [0,T]$, on the lattice $k\Delta t, k \in \mathbb{N}_n$ is defined as follows:
\begin{align}
	f_{k+1,n} &= \begin{cases}
	f_{k+1,n}^{(u)} ,& \textrm{if} \; \zeta^{(p_{\Delta t})}_{k+1,n} = 1 \\
	f_{k+1,n}^{(d)} ,& \textrm{if} \; \zeta^{(p_{\Delta t})}_{k+1,n} = 0 
	\end{cases},k = 0,...,n-1,
	\label{eq_payoff13}
\end{align}
and $f_{n,n} = g(S_T),f_{k,n} = f(S_{k \Delta t, k \Delta t}),k=0,...,n$. At time instances $k\Delta t, k = 0,...,n-1$, trader $\aleph$, taking a short position in $\mathcal{C}$, is forming a self-financing replicating risk-neutral portfolio $P_{k\Delta t;n} = \mathbb{D}_{k\Delta t}S_{k,n}^{(p_{\Delta t})}-f_{k,n}$. Conditionally on $\mathcal{F}_{k,n},\;P_{(k+1)\Delta t;n} = \mathbb{D}_{k \Delta t} S^{(p_{\Delta t})}_{(k+1),n}-f_{k+1,n}$. As demonstrated by KSRF, the risk-neutrality condition implies  that, conditionally on $\mathcal{F}_{k;n}$,
\begin{equation}
	f_{k,n} = e^{-r \Delta t}(q_{\Delta t}^{(p_{\Delta t})} f_{k+1,n}^{(u)}+(1-q_{\Delta t}^{(p_{\Delta t})})f_{k+1,n}^{(d)}), k = 0,...,n-1.
	\label{eq_payoff14}
\end{equation}
The risk-neutral probability $q_{\Delta t}$ in (\ref{eq_payoff14}) is given by
\begin{equation}
	q_{\Delta t}^{(p_{\Delta t})} = \frac{\exp \left\{-(r-\mu)\Delta t \right\}-\exp \left\{-\frac{p_{\Delta t}}{1-p_{\Delta t}} \frac{\sigma^2 \Delta t}{2} - \sigma \sqrt{\frac{p_{\Delta t}}{1-p_{\Delta t}}\Delta t}\right\}}
	{\exp \left\{-\frac{1-p_{\Delta t}}{p_{\Delta t}} \frac{\sigma^2 \Delta t}{2} - \sigma \sqrt{\frac{1-p_{\Delta t}}{p_{\Delta t}}\Delta t}\right\}-\exp \left\{- \frac{p_{\Delta t}}{1-p_{\Delta t}} \frac{\sigma^2 \Delta t}{2} - \sigma \sqrt{\frac{p_{\Delta t}}{1-p_{\Delta t}}\Delta t}\right\}}.	
	\label{eq_tree_q_15}
\end{equation}
With $o(\Delta t) = 0$, $q_{\Delta t}$ has the form
\begin{equation}
	q_{\Delta t}^{(p_{\Delta t})} = p_{\Delta t}- \theta \sqrt{p_{\Delta t}(1-p_{\Delta t})}\sqrt{\Delta t},
	\label{eq_tree_q_16}
\end{equation}
where $\theta = \frac{\mu-r}{\sigma}$ is the market price of risk. The risk-neutral pricing tree is given by
\begin{equation}
	S_{k+1,n}^{(q_{\Delta t})} = 
	\begin{cases} 
      S_{k+1,n}^{(q_{\Delta t},u)} = S_{k,n}^{(q_{\Delta t})}e^{U_{\Delta t}},& \textrm{if} \; \zeta_{k+1,n}^{(q_{\Delta t})} = 1 \\
      S_{k+1,n}^{(q_{\Delta t},d)} = S_{k,n}^{(q_{\Delta t})}e^{D_{\Delta t}},& \textrm{if} \; \zeta_{k+1,n}^{(q_{\Delta t})} = 0
	\end{cases}
	= S_{k,n}^{(q_{\Delta t})} 
	\begin{cases} 
      e^{U_{\Delta t}}, & w.p.\; q_{\Delta t},\\
      e^{D_{\Delta t}}, & w.p.\; 1-q_{\Delta t}.
	\end{cases}
	\label{eq_tree17}
\end{equation}
By the DPIP, the $\mathcal{D}[0,T]$-process, 
\begin{align*}
\mathbb{S}^{(n;\mathbb{Q})} = \{S_t^{(n;\mathbb{Q})} = S_{k,n}^{(q_{\Delta t})},t\in [k\Delta t,(k+1)\Delta t),k=0,1,...,n-1,S_T^{(n;\mathbb{Q})}=S_{n,n}^{(q_{\Delta t})}\},
\end{align*}
 converges weakly in $\mathcal{D}[0,T]$ to $\mathbb{S}^{(n;\mathbb{Q})}=\{S_t^{(\mathbb{Q})} = S_t^{(\mathbb{Q};r,\sigma)},t \in [0,T]\}$, where $S_t^{(\mathbb{Q})} = S_t^{(\mathbb{Q};r,\sigma)} = S_0e^{(r-\frac{1}{2}\sigma^2)t+\sigma B_t^{(\mathbb{Q})}}$, and $B_t^{(\mathbb{Q})},t\in [0,T]$, is a BM on $(\Omega,\mathbb{F} = \{\mathcal{F}_t  \}_{t \geq 0},\mathbb{Q})$. The probability measure $\mathbb{Q}\sim\mathbb{P}$ is the unique equivalent martingale measure.\footnote{See Chapter 6 in Duffie (2001).} The limiting continuous-time price process $S_t^{(\mathbb{Q})}, t \in [0,T]$ is now independent of $p_{\Delta t}$ and $\mu$. This is due to the assumption that $\aleph$ can hedge his short position in the option contract continuously in time. However, if $\aleph$'s hedging trading times  are restricted to the time instances $k\Delta t,k = 0,1,...,n-1$, the pricing tree (\ref{eq_tree17}) does depend on $p_{\Delta t}$ and $\mu$, due to (\ref{eq_payoff14}) and (\ref{eq_tree_q_15}). Furthermore, the discontinuity of the option price at $p_{\Delta t}\rightarrow 0$, or $p_{\Delta t}\rightarrow 1$, does not exist anymore, because according to (\ref{eq_tree_q_16}), $\lim_{p_{\Delta t} \uparrow 1} q_{\Delta t} = 1$, and $\lim_{p_{\Delta t} \downarrow 0} q_{\Delta t} = 0$.

\subsection{Implied $\mu$-surface and implied $p_{\Delta t}$-surface}
In the previous section we showed the dependence of risk-neutral pricing tree (\ref{eq_tree17}) on stock mean return $\mu$ and the probability for stock upturn $p_{\Delta t}$. Thus, similarly to the concept of implied volatility, we introduce the concept of \textit{implied $\mu$-surface} and \textit{implied $p_{\Delta t}$-surface}, illustrated in the following numerical example.

The framework is based on KSRF binomial option pricing tree of (\ref{eq_tree12}), (\ref{eq_payoff14}), and (\ref{eq_tree_q_16}) in Section 2.3. In the simulation, we use daily trading frequencies of SPDR S\&P 500 ETF(SPY)\footnote{\url{https://finance.yahoo.com/quote/SPY? p=SPY\& .tsrc=fin-srch.}} and  corresponding Mini-SPX(XSP)\footnote{The CBOE Mini-SPX (with ticker XSP) option contract is an index option product designed to track the underlying S\&P 500 Index with the size of 1/10 of the standard SPX options contract. See \url{http://www.cboe.com\iffalse /products/stock-index-options-spx-rut-msci-ftse/s-p-500-index-options/mini-spx-index-options-xsp\fi.}} call option prices as datasets. We use SPY to estimate the initial $\hat{p}$, $\hat{\mu}$, and $\hat{\sigma}$ using one-year back trading data with the time range from $5/16/2019$ to $5/15/2020$. And we use the 10-year Treasury yield curve rate\footnote{\url{https://www.treasury.gov\iffalse /resource-center/data-chart-center/interest-rates/pages/textview.aspx?data=yield\fi.}} of the starting date as the riskless rate $r$. To estimate $\hat{p}$ which is the probability of the sample price increasing for a fixed day, we use the proportion of the number of days with non-negative log-return in one-year back trading period. We set $\hat{\sigma}$ as the sample standard deviation of sample return series, and $\Delta t = \frac{1}{252}$.

The starting date for the option is $5/15/2020$ with different call option contracts varying from $5/18/2020$ to $12/07/2021$. And we have the closing price of SPY at $5/15/2020$ as $S_0 =\$ 286.28$ with $r = 0.64\%$. Then, the initial values for the parameters can be estimated: $\hat{p}_0 = 0.56$, $\hat{\mu}_0 = 1.80\times 10^{-4}$, and $\hat{\sigma}_0 = 0.02$.

To get the implied $\mu$-surface, we set $\mu$ as a free parameter, and then match the XSP option price $C^{(market;\mathcal{C})}(S_0,K,T,r,\hat{p},\hat{\sigma})$ with the theoretical call option derived by (\ref{eq_payoff14}). For the $i^{th}$ XSP contract in the sample, we have
\begin{align*}
	\mu^{(\aleph;impled,i)} = \textup{arg\,min}\left(\frac{C^{(\aleph;\mathcal{C},i)}(S_0,K,T,r,\hat{p},\hat{\sigma})-C^{(market;\mathcal{C},i)}(S_0,K,T,r,\hat{p},\hat{\sigma})}{C^{(market;\mathcal{C},i)}(S_0,K,T,r,\hat{p},\hat{\sigma})}\right)^2.
\end{align*}
Similarly, we switch the free parameter from $\mu$ to $p$ to get the implied $p$-surface. 

\begin{figure}[htb!]
	\centering
	\includegraphics[scale=0.32,angle=0]{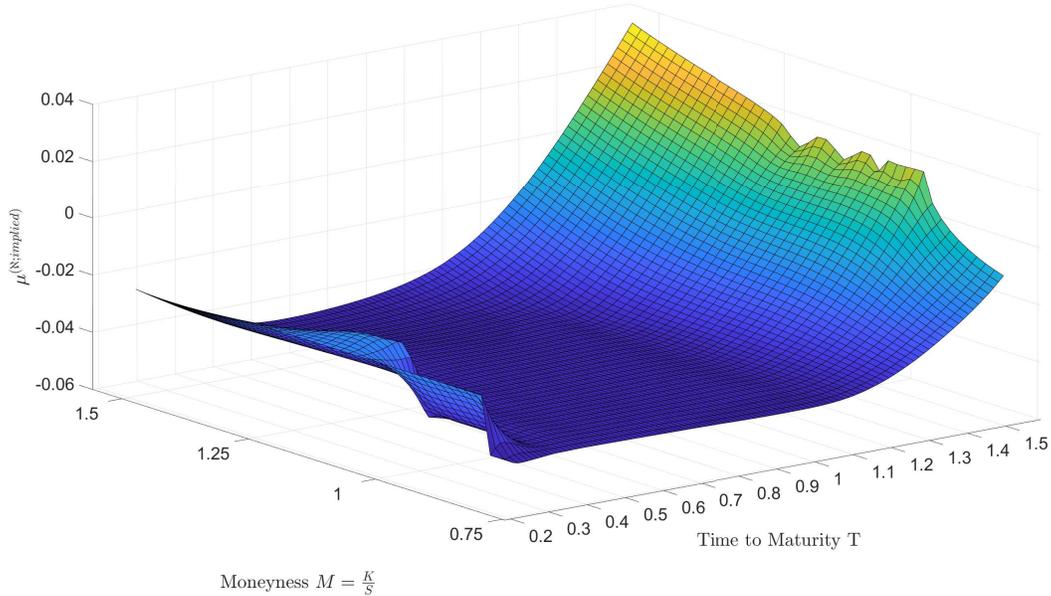}
	\caption{Implied $\mu$--surface against time to maturity and moneyness.}
	\label{fig:task1_mu}
\end{figure}

As shown in Figure \ref{fig:task1_mu}, the implied $\mu$-surface is against “moneyness” ($M$)\footnote{Here, we define moneyness $ M = \frac{K}{S}$, where $K$ is the strike and $S$ is the price.} and time to maturity $T$ in years. According to Figure \ref{fig:task1_mu}, the $\mu^{(\aleph;implied)}\in(-0.04,0.04)$. For a fixed $M = 1.5$, for example, the $\mu^{(\aleph;implied)}$ decreases from $-0.02$ to $-0.04$ for about three months and is stable for another eight months, then it recovers sharply to $0.04$. At certain maturity time $t\in[0,T],T = 1.5$, Figure\ref{fig:task1_mu} indicates that the $\mu^{(\aleph;implied)}$ for the option trader slightly increases as $M$ increases. And the increment is easier to capture after one year.

\begin{figure}[htb!]
	\centering
	\includegraphics[scale=0.32,angle=0]{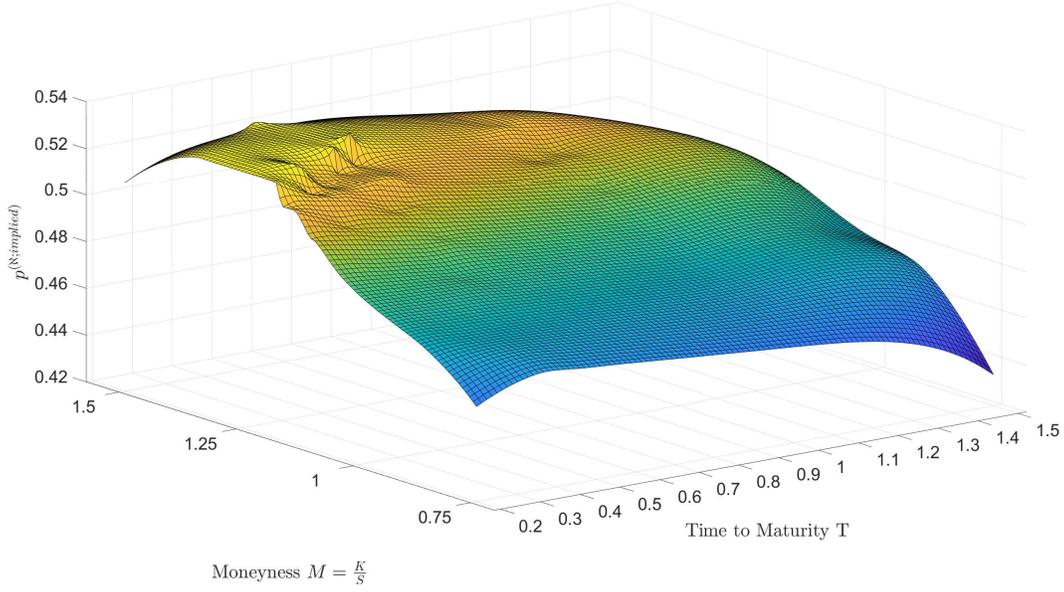}
	\caption{Implied $p_{\Delta t}$--surface against time to maturity and moneyness.}
	\label{fig:task1_p}
\end{figure}

Figure \ref{fig:task1_p} shows the result of $p_{\Delta t}^{(\aleph;implied)}$ which is the implied probability according to KSRF binomial option pricing. Similarly, the implied $p_{\Delta t}$-surface is plotted against $M$ and $T$. In Section 2.2, we have that $\aleph$ is an \textit{informed trader} if $p_{\Delta t}^{\aleph} > \frac{1}{2}$. More specifically, option traders are potentially more informed about future SPY returns better than spot traders. On the other hand, $\aleph$ is a \textit{misinformed trader} if $p_{\Delta t}^{\aleph} < \frac{1}{2}$, which describes the situation of spot traders more aware of the future movement of SPY than option traders. Our result indicates $p_{\Delta t}^{(\aleph;implied)}\in(0.47,0.53)$. For a fixed $M$, the implied probability of option traders decreases as $T$ increases. This fact indicates that option traders are informed in the near future rather than in the distance future. For a fixed $T$, $p_{\Delta t}^{(\aleph;implied)}$ is roughly greater than $0.5$ when $M\in(1,1.5)$ and less than $0.5$ when $M\in(0.75,1)$. This indicates that option traders are more informed when $M$ is high than when $M$ is small according to Figure \ref{fig:task1_p}.
\subsection{Information distance as a measure of an option trader's information on stock price direction}
\par
\begin{figure}[htb!]
	\centering
	\includegraphics[scale=0.32,angle=0]{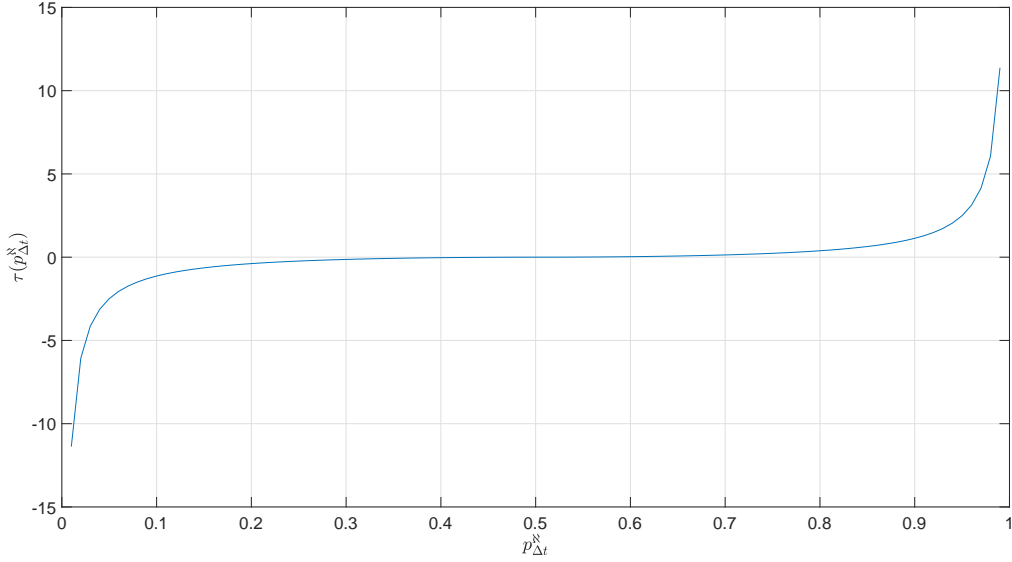}
	\caption{Information level $\tau(p_{\Delta t}^{\aleph})$, where $0<p_{\Delta t}^{\aleph}<1$. }
	\label{fig:tau_p}
\end{figure}

{\fontsize{10}{13}To quantify the amount of information of an informed trader $\aleph$ with $p_{\Delta t}^{\aleph}>\frac{1}{2}$, we use Shannon's entropy\footnote{See, for example, Robinson (2008) and Rioul (2018).} as an information measure. Shannon's entropy of a Bernoulli random variable $\zeta^{(p)} \overset{\mathrm{d}}{=} Ber(p), \mathbb{P}(\zeta^{(p)} = 1) = 1-\mathbb{P}(\zeta^{(p)} = 0) = p \in (0,1)$ is defined as
\begin{equation}
	H(\zeta^{(p)}) = -p \ln p-(1-p) \ln(1-p),
\end{equation}
and $\max_{0<p<1}H(\zeta^{(p)}) = H(\zeta^{(\frac{1}{2})}) = \log 2$.\footnote{See Chapter 1 in Cover \& Thomas (2006) and Chapter 2 in Billinglsey (1965).} The information distance (the relative entropy, the Kullback-Leibler divergence\footnote{See Chapter 2 in Cover \& Thomas (2006) and Rioul (2018).}) between $\zeta^{(p)}$ and $\zeta^{(\frac{1}{2})}$ is determined by
\begin{equation}
	D(\zeta^{(p)},\zeta^{(\frac{1}{2})}) = p\log(2p)+(1-p)\log (2-2p).
\end{equation}
As $\max_{0<p<1}D(\zeta^{(p)},\zeta^{(\frac{1}{2})}) = \lim_{p\uparrow 1}D(\zeta^{(p)},\zeta^{(\frac{1}{2})}) = \lim_{p\downarrow 0}D(\zeta^{(p)},\zeta^{(\frac{1}{2})}) = \log 2$, we define the $\aleph$'s information level as
\begin{equation}
	\tau(p_{\Delta t}^{\aleph}) = \textup{sign}\left( p_{\Delta t}^{\aleph}-\frac{1}{2}\right)\frac{D(\zeta^{(p_{\Delta t}^{\aleph})},\zeta^{(\frac{1}{2})})}{\log 2 - D(\zeta^{(p_{\Delta t}^{\aleph})},\zeta^{(\frac{1}{2})})},p_{\Delta t}^{\aleph}\in (0,1).
\end{equation}}
Figure \ref{fig:tau_p} shows the trend of $\tau(p_{\Delta t}^{\aleph})$ as $p_{\Delta t}^{\aleph}\in (0,1)$, where the value of $\tau(p_{\Delta t}^{\aleph})$ increases when $p_{\Delta t}^{\aleph}$ increases.
\section{Option Trading When The Hedger Has Information on Stock Price Direction}
In this section we address the question of $\aleph$'s potential gain from trading with the information level $\tau(p_{\Delta t}^{\aleph})>0$.\footnote{According to the Efficient Market Hypothesis (EMH), asset price direction is unpredictable, see Fama (1970). However, some studies indicate that asset price  direction is predictable  (and, thus, questioning the EMH). See, among others, Shiller (2003 and 2013).} We assume that $\aleph$ knows, with probability $p_{\Delta t}^{\aleph}>\frac{1}{2}$,\footnote{We assume that $\aleph$ is an informed trader, that is, $p_{\Delta t}^{\aleph}>\frac{1}{2}$. We will develop a trading strategy for $\aleph$ to utilize his information on stock price direction. A misinformed trader will trade just the opposite of what an informed trader will do, and what will be a profit for the informed trader will be a loss for the misinformed trader. Thus, it is sufficient to consider the case of $\aleph$  being an informed trader. We shall summarize the results for informed and misinformed traders at the end of Section 3.2.} the stock price direction in period $[k \Delta t, (k+1)\Delta t]$. We also assume that in the marketplace, there are a sufficient number of noisy  traders, $\aleph_0$, whose trading activities are based on the assumption that $p_{\Delta t} = \frac{1}{2}$ in (\ref{eq_option10}). At any time instance, $k \Delta t, k=0,...,n,n\Delta t = T$, $\aleph$ makes independent bets, which are modeled as independent Bernoulli trials $\eta_{k+1,n}^{(\aleph)},k=0,...,n-1,\mathbb{P}(\eta_{k+1,n}^{(\aleph)} = 1)=1-\mathbb{P}(\eta_{k+1,n}^{(\aleph)} = 0)= p_{\Delta t}^{\aleph}\in (\frac{1}{2},1)$. We consider the following up- and 
down-scenarios: $(Sc^{(up)})\zeta_{k+1,n}^{(p_{\Delta t})} = 1$; that is, $S_{k+1,n}^{(p_{\Delta t})} = S_{k+1,n}^{(p_{\Delta t},u)} = S_{k,n}^{(p_{\Delta t})}e^{U_{\Delta t}}$, and $(Sc^{(down)})\zeta_{k+1,n}^{(p_{\Delta t})} = 0$, that is, $S_{k+1,n}^{(p_{\Delta t})} = S_{k+1,n}^{(p_{\Delta t},d)} = S_{k,n}^{(p_{\Delta t})}e^{D_{\Delta t}}$. Now the filtration $\mathbb{F}^{(n)} = \{\mathcal{F}_{k;n} = \sigma(\zeta_{1,n}^{(p_{\Delta t})},...,\zeta_{k,n}^{(p_{\Delta t})}),k \in \mathbb{N}_n,\mathcal{F}_{0;n} = \{\varnothing, \Omega\}\}$ needs to be augmented with the sequence of $\aleph$'s independent bets. We introduce the augmented filtration $\mathbb{F}^{(n;\aleph)} = \{\mathcal{F}_{k;n}^{\aleph} = \sigma((\zeta_{1,n}^{(p_{\Delta t})},\eta_{1,n}^{(\aleph)}),...,(\zeta_{k,n}^{(p_{\Delta t})},\eta_{k,n}^{(\aleph)})),k \in \mathbb{N}_n,\mathcal{F}^{(\aleph)}_{0;n} = \{\varnothing, \Omega\}\}$. 

\subsection{Forward contract’s strategy for a trader with information on stock price direction}
At $k\Delta t, k = 0,...,n-1$, $\aleph$ places his bets considering $(Sc^{(up)})$ and $(Sc^{(down)})$. If at $k\Delta t$, $\aleph$ believes that $(Sc^{(up)})$ will happen, he takes a long position\footnote{The short position in the forward contract could be taken by any trader who believes that $S_{k+1,n}^{(p_{\Delta t})} = S_{k+1,n}^{(p_{\Delta t},d)} = S_{k,n}^{(p_{\Delta t})}e^{D_{\Delta t}}$ is more likely to happen, or by a noisy trader $\aleph_0$.} in $\Delta ^{(\aleph)}_{k \Delta t} = \frac{N^{(\aleph)}}{S_{k,n}^{(p_{\Delta t})}}$-forward contracts for some $N^{(\aleph)}>0$.\footnote{Parameter $N^{(\aleph)}$ will be optimized and will enter the formula for the positive yield $\aleph$ will enjoy when trading options, see Section 3.2.} The maturity of the forwards is $(k+1)\Delta t$. If at $k \Delta t$, $\aleph$ believes that $(Sc^{(down)})$ will happen, he takes a short position in $\Delta ^{(\aleph)}_{k \Delta t}$-forward contracts\footnote{The long position in the forward contract could be taken by any trader who believes that $S_{k+1,n}^{(p_{\Delta t})} = S_{k+1,n}^{(p_{\Delta t},u)} = S_{k,n}^{(p_{\Delta t})}e^{U_{\Delta t}}$ is more likely to happen, or by a noisy trader $\aleph_0$.} at maturity $(k+1)\Delta t$. The overall payoff of  $\aleph$'s forward contract positions is given by
\begin{equation}
	p_{k \Delta t \rightarrow (k+1) \Delta t}^{(\aleph;forward)} = \Delta_{k \Delta t}^{(\aleph)}
	\begin{cases} 
      (S_{k+1,n}^{(p_{\Delta t},u)}-S_{k,n}^{(p_{\Delta t})}e^{r\Delta t}),& \textrm{if} \; \zeta_{k+1,n}^{(p_{\Delta t})} = 1, \eta_{k+1,n}^{(\aleph)} = 1,
       \\
      (S_{k,n}^{(p_{\Delta t})}e^{r\Delta t}-S_{k+1,n}^{(p_{\Delta t},d)}),& \textrm{if} \; \zeta_{k+1,n}^{(p_{\Delta t})} = 0, \eta_{k+1,n}^{(\aleph)} = 1,
      \\
      (S_{k,n}^{(p_{\Delta t})}e^{r\Delta t}-S_{k+1,n}^{(p_{\Delta t},u)}),& \textrm{if} \; \zeta_{k+1,n}^{(p_{\Delta t})} = 1, \eta_{k+1,n}^{(\aleph)} = 0,
      \\
      (S_{k+1,n}^{(p_{\Delta t},d)}-S_{k,n}^{(p_{\Delta t})}e^{r\Delta t}),& \textrm{if} \; \zeta_{k+1,n}^{(p_{\Delta t})} = 0, \eta_{k+1,n}^{(\aleph)} = 0.
	\end{cases}
\end{equation}

The conditional mean and variance of $p_{k \Delta t \rightarrow (k+1) \Delta t}^{(\aleph;forward)}$ are given by
\begin{align}
	\mathbb{E}(p_{k \Delta t \rightarrow (k+1) \Delta t}^{(\aleph;forward)}| \mathcal{F}^{(\aleph)}_{k;n}) &= N^{(\aleph)}(2p_{\Delta t}^{\aleph}-1)\sigma\left(\theta(2p_{\Delta t}-1)\Delta t+2\sqrt{p_{\Delta t}(1-p_{\Delta t})\Delta t}\right), \nonumber 
	\\
	Var(p_{k \Delta t \rightarrow (k+1) \Delta t}^{(\aleph;forward)}| \mathcal{F}^{(\aleph)}_{k;n}) &=	N^{(\aleph)^2}\sigma^2\left(1-4(2p_{\Delta t}^{\aleph}-1)^2p_{\Delta t}(1-p_{\Delta t})\right)\Delta t.
\label{eq_meanvar_22}
\end{align}
where $\theta = \frac{\mu-r}{\sigma}$. By the DPIP, we should have $\mathbb{E}(p_{k \Delta t \rightarrow (k+1) \Delta t}^{(\aleph;forward)}| \mathcal{F}^{(\aleph)}_{k;n}) = O(\Delta t)$ and $Var(p_{k \Delta t \rightarrow (k+1) \Delta t}^{(\aleph;forward)}| \mathcal{F}^{(\aleph)}_{k;n}) = O(\Delta t)$. To guarantee that, we set $p^{(\aleph)}_{\Delta t} = \frac{1}{2}(1+\frac{\lambda^{(\aleph)}}{\sqrt{p_{\Delta t}(1-p_{\Delta t})}}\sqrt{\Delta t})$, for some $\lambda^{(\aleph)}>0$.\footnote{The case of a misinformed trader can be considered in a similar manner. A misinformed trader with $\lambda^{(\aleph)}<0$, trades long-forward (resp. short-forward) when the informed trader with $(-\lambda^{(\aleph)})>0$, trades short-forward (resp. long-forward). A noisy trader will not trade any forward contracts, as he has no information about stock price direction.} The closer $p_{\Delta t}$ is to $1$, or $0$, the more certain will be $\aleph$ on stock price direction, and thus $p^{(\aleph)}_{\Delta t}$ increases.
Then, (\ref{eq_meanvar_22}) simplifies to
\begin{align}
	\mathbb{E}(p_{k \Delta t \rightarrow (k+1) \Delta t}^{(\aleph;forward)}| \mathcal{F}^{(\aleph)}_{k;n}) &= 2N^{(\aleph)}\lambda^{(\aleph)}\sigma \Delta t, \nonumber
	\\
	Var(p_{k \Delta t \rightarrow (k+1) \Delta t}^{(\aleph;forward)}| \mathcal{F}^{(\aleph)}_{k;n}) &=	N^{(\aleph)^2}\sigma^2 \Delta t.
\label{eq_meanvar_23}
\end{align}
The \textit{instantaneous information ratio}\footnote{We have chosen the  normalization $\frac{1}{p_{\Delta t}(1-p_{\Delta t})}$ for $\lambda^{(\aleph)}$ in $p^{(\aleph)}_{\Delta t} = \frac{1}{2}(1+\frac{\lambda^{(\aleph)}}{\sqrt{p_{\Delta t}(1-p_{\Delta t})}}\sqrt{\Delta t})$, so that $IR(p_{k \Delta t \rightarrow (k+1) \Delta t}^{(\aleph;forward)}| \mathcal{F}^{(\aleph)}_{k;n}) = 2\lambda^{(\aleph)}$ is solely dependent on $\lambda^{(\aleph)}$.} is given by
\begin{equation}
	IR(p_{k \Delta t \rightarrow (k+1) \Delta t}^{(\aleph;forward)}| \mathcal{F}^{(\aleph)}_{k;n}) = \frac{\mathbb{E}(p_{k \Delta t \rightarrow (k+1) \Delta t}^{(\aleph;forward)}| \mathcal{F}^{(\aleph)}_{k;n})}{\sqrt{\Delta t}\sqrt{Var(p_{k \Delta t \rightarrow (k+1) \Delta t}^{(\aleph;forward)}| \mathcal{F}^{(\aleph)}_{k;n})}} = 2\lambda^{(\aleph)}.
\end{equation}

\subsection{Option pricing for trader with information on the stock price direction}
Suppose now that $\aleph$ is taking a short position in the option contract within the BSM framework $(\mathcal{S},\mathcal{B},\mathcal{C})$\footnote{The long position in the option contract is taken by a trader who trades the stock with stock dynamics given by (\ref{eq_price1}).}. The stock price dynamics $S_t = S_t^{(\mu,\sigma)},t \geq 0$, is given by (\ref{eq_price1}), the bond price $\beta_t, t \geq 0$ is given by (\ref{eq_beta2}), and the derivative $\mathcal{C}$ has price $f_t = f(S_t,t),t \in [0,T]$ with terminal payoff $f_T = g(S_T)$. When $\aleph$ trades the stock $\mathcal{S}$, hedging the short position in $\mathcal{C}$, $\aleph$ simultaneously runs his forward strategy. $\aleph$'s trading strategy (a combination of the forward contact's trading and  trading the stock)  leads to an enhanced price process, of which dynamics can be expressed as follows: $S_{0,n}^{(\aleph;\mathcal{C})} = S_0$ and
\begin{equation}
	S_{k+1,n}^{(\aleph;\mathcal{C})} = 
	\begin{cases} 
      S_{k+1,n}^{(p_{\Delta t},u)}+N^{(\aleph)}(S_{k+1,n}^{(p_{\Delta t},u)}-S_{k,n}^{(p_{\Delta t})}e^{r\Delta t}),\; \textrm{if} \; \zeta_{k+1,n}^{(p_{\Delta t})} = 1, \eta_{k+1,n}^{(\aleph)} = 1, \\
      S_{k+1,n}^{(p_{\Delta t},d)}+N^{(\aleph)}(S_{k,n}^{(p_{\Delta t})}e^{r\Delta t}-S_{k+1,n}^{(p_{\Delta t},d)}),\; \textrm{if} \; \zeta_{k+1,n}^{(p_{\Delta t})} = 0, \eta_{k+1,n}^{(\aleph)} = 1, \\
      S_{k+1,n}^{(p_{\Delta t},u)}+N^{(\aleph)}(S_{k,n}^{(p_{\Delta t})}e^{r\Delta t}-S_{k+1,n}^{(p_{\Delta t},u)}),\; \textrm{if} \; \zeta_{k+1,n}^{(p_{\Delta t})} = 1, \eta_{k+1,n}^{(\aleph)} = 0, \\
      S_{k+1,n}^{(p_{\Delta t},d)}+N^{(\aleph)}(S_{k+1,n}^{(p_{\Delta t},d)}-S_{k,n}^{(p_{\Delta t})}e^{r\Delta t}),\; \textrm{if} \; \zeta_{k+1,n}^{(p_{\Delta t})} = 0, \eta_{k+1,n}^{(\aleph)} = 0,
	\end{cases}
	\label{eq_op_price25}
\end{equation}
$k = 0,1,...,n-1,n\Delta t = T.$\footnote{With every single share of the traded stock with price  $S_{k,n}^{(p_\Delta t ) }$  at $k\Delta t$, $\aleph$ simultaneously enters $N^{(\aleph)}$-forward contracts. The forward contracts are long or short, depending on $\aleph$'s views on stock price direction in time-period $[k\Delta t,(k+1)\Delta t]$.} It costs nothing to enter a forward contract at $k\Delta t$ with terminal time $(k+1)\Delta t$. Then,
\begin{align*}
	\mathbb{E}(\frac{S_{k+1,n}^{(\aleph;\mathcal{C})}}{S_{k,n}^{(\aleph;\mathcal{C})}}|S_{k,n}^{(\aleph;\mathcal{C})}) 
	&= 1+\left(\mu+N^{(\aleph)}(\mu-r)(2p_{\Delta t}-1)(2p^{\aleph}_{\Delta t}-1)\right)\Delta t 
	\\
	&+2N^{(\aleph)}\sigma \sqrt{(1-p_{\Delta t})p_{\Delta t}}(2p^{\aleph}_{\Delta t}-1)\sqrt{\Delta t}.
\end{align*}
As already discussed in Section 3.1, we set $2p^{(\aleph)}_{\Delta t}-1 = \frac{\lambda^{(\aleph)}}{\sqrt{p_{\Delta t}(1-p_{\Delta t})}}\sqrt{\Delta t}, \lambda^{(\aleph)}>0$. Thus, the conditional mean and variance of the log-return $R_{k,n}^{(\aleph;\mathcal{C})}=\log(\frac{S_{k+1,n}^{(\aleph;\mathcal{C})} }{S_{k,n}^{(\aleph;\mathcal{C})} })$ are $\mathbb{E}(R_{k,n}^{(\aleph;\mathcal{C})} |S_{k,n}^{(\aleph;\mathcal{C})} ) = (\mu+2N^{(\aleph)}\sigma \lambda^{(\aleph)})\Delta t$ and $Var(R_{k,n}^{(\aleph;\mathcal{C})} |S_{k,n}^{(\aleph;\mathcal{C})} ) = \sigma^2(1+N^{(\aleph)^2})\Delta t$. The instantaneous market price of risk is given by 
\begin{equation}
	\Theta(R_{k,n}^{(\aleph;\mathcal{C})} |S_{k,n}^{(\aleph;\mathcal{C})} ) = \frac{\mathbb{E}(R_{k,n}^{(\aleph;\mathcal{C})} |S_{k,n}^{(\aleph;\mathcal{C})}) - r\Delta t}{\sqrt{\Delta t}\sqrt{Var(R_{k,n}^{(\aleph;\mathcal{C})} |S_{k,n}^{(\aleph;\mathcal{C})} )}} = \frac{\theta+2N^{(\aleph)} \lambda^{(\aleph)}}{\sqrt{1+N^{(\aleph)^2}}}.
	\label{eq_op_price27}
\end{equation}
The optimal $N^{(\aleph)}$, maximizing $\Theta(R_{k,n}^{(\aleph;\mathcal{C})} |S_{k,n}^{(\aleph;\mathcal{C})} )$, is $N^{(\aleph)} = N^{(\aleph;opt)} = 2\frac{\lambda^{(\aleph)}}{\theta}>0$,\footnote{By assumption, $\mu>r>0$, and thus, $\theta = \frac{\mu-r}{\sigma}>0$.} and the optimal instantaneous market price of risk is   $\Theta(R_{k,n}^{(\aleph;\mathcal{C})} |S_{k,n}^{(\aleph;\mathcal{C})} ) = \Theta^{(opt)}(R_{k,n}^{(\aleph;\mathcal{C})} |S_{k,n}^{(\aleph;\mathcal{C})} ) = \sqrt{\theta^2+4\lambda^{{(\aleph)}^2}}$. With $N^{(\aleph)} = N^{(\aleph;opt)}$,
\begin{align}
	\mathbb{E}(R_{k,n}^{(\aleph;\mathcal{C})} |S_{k,n}^{(\aleph;\mathcal{C})}) &= (\mu+4\sigma\frac{\lambda^{{(\aleph)}^2}}{\theta})\Delta t, \nonumber
	\\
	Var(R_{k,n}^{(\aleph;\mathcal{C})} |S_{k,n}^{(\aleph;\mathcal{C})}) &= \sigma^2(1+4\frac{\lambda^{{(\aleph)}^2}}{\theta^2})\Delta t.
\label{eq_meanvar29}
\end{align}

Next, we consider the limiting behavior of $\mathbb{S}^{(n,\aleph;\mathcal{C})} = \{S_t^{(n,\aleph;\mathcal{C})} = S_{k,n}^{(\aleph;\mathcal{C})},t\in [k \Delta t, (k+1) \Delta t),k =0,...,n, S_T^{(n,\aleph;\mathcal{C})} = S_{n,n}^{(\aleph;\mathcal{C})}\}$ as $\Delta t \downarrow 0$. We set $p_{\Delta t} = p_0 \in (0,1)$. By the DPIP and (\ref{eq_op_price25}), it follows that, $\mathbb{S}^{(n,\aleph;\mathcal{C})}$ converges weakly in $\mathcal{D}[0,T]$ to $\mathbb{S}^{(\aleph;\mathcal{C})} = \{S_t^{(\aleph;\mathcal{C})},t \in [0,T]\}$ as $n \uparrow \infty$, where
\begin{equation}
	S_t^{(\aleph;\mathcal{C})} = S_0 \exp{\left\{(\mu^{(\aleph;\mathcal{C})}-\frac{1}{2}\sigma^{(\aleph;\mathcal{C})^2})t+\sigma^{(\aleph;\mathcal{C})}B_t\right\}}.
\end{equation}
In (\ref{eq_meanvar29}), $\mu^{(\aleph;\mathcal{C})} = \mu+4\sigma\frac{\lambda^{{(\aleph)}^2}}{\theta}$, and $\sigma^{(\aleph;\mathcal{C})} = \sigma\sqrt{1+4\frac{\lambda^{{(\aleph)}^2}}{\theta^2}}$. Now, in the limit $\Delta t \downarrow 0$, $\aleph$ hedges the short option position using the price process $\mathbb{S}^{(\aleph;\mathcal{C})}$. $\aleph$ forms his instantaneous riskless replicating portfolio $\Pi^{(\aleph;\mathcal{C})}_t =a_t^{(\aleph;\mathcal{C})} S_t^{(\aleph;\mathcal{C})}+b_t^{(\aleph;\mathcal{C})} \beta_t = f_t,t\in[0,T) $. As $\Pi^{(\aleph;\mathcal{C})}_t$ is self-financing portfolio, and thus, $df_t = d\Pi^{(\aleph;\mathcal{C})}_t =a_t^{(\aleph;\mathcal{C})} dS_t^{(\aleph;\mathcal{C})}+b_t^{(\aleph;\mathcal{C})} d\beta_t$. By It$\hat{\textrm{o}}$'s formula,
\begin{align}
&\left(\frac{\partial f(S_t,t)}{\partial t}+\mu S_t \frac{\partial f(S_t,t)}{\partial x}+\frac{1}{2}\sigma^2 S_t^2 \frac{\partial^2 f(S_t,t)}{\partial x^2}\right)dt+\sigma S_t\frac{\partial f(S_t,t)}{\partial x}dB_t \nonumber
\\
&= a_t^{(\aleph;\mathcal{C})} S_t^{(\aleph;\mathcal{C})}(\mu^{(\aleph;\mathcal{C})}dt+\sigma^{(\aleph;\mathcal{F})}dB_t)+b_t^{(\aleph;\mathcal{C})} r\beta_tdt. \label{eq_partial30}
\end{align}
Because the forward contract, which $\aleph$ initiates at $t$, has zero value, then $S_t^{(\aleph;\mathcal{C})} = S_t$ in (\ref{eq_partial30}). The no-arbitrage argument implies that $a_t^{(\aleph;\mathcal{C})} = \frac{\partial f(S_t,t)}{\partial x}\frac{\sigma}{\sigma^{(\aleph;\mathcal{C})}}$ and $b_t^{(\aleph;\mathcal{C})} = \frac{1}{\beta_t}\left(f(S_t,t) - a_t^{(\aleph;\mathcal{C})}S_t^{(\aleph;\mathcal{C})}\right)$. Thus,  the BSM partial differential equation (PDE) for $\aleph$'s option price $f_t = f(x,t),x>0,t\in [0,T)$, is given by
\begin{equation}
	\frac{\partial f(x,t)}{\partial t}+(r-D_y)x \frac{\partial f(x,t)}{\partial x} +\frac{1}{2}\sigma^2 x^2 \frac{\partial^2 f(x,t)}{\partial x^2} -rf(x,t) = 0.
\label{eq_op31}
\end{equation}
The continuous dividend yield $D_y \in R$ has the form  $D_y = D_y ^{(\aleph;\mathcal{C})}  = \sigma(\sqrt{\theta^2+4\lambda^{(\aleph)^2}}-\theta)>0$. According to (\ref{eq_op31}), the BSM formula for the European call option price for an informed trader $\aleph$, $C_t = C(S_t,t),t\in [0,T]$ with strike price $K$,  is given by the standard option-price formula for the dividend-paying stock:
\begin{align}
	C^{(\aleph;\mathcal{C})}(S_t,K,T-t,r,\sigma,D_y) &= e^{-D_y(T-t)}N(d_1)S_t-N(d_2)Ke^{-r(T-t)},
	\nonumber	
	\\
	d_1 &= \frac{1}{\sigma \sqrt{T-t}}\left[\log(\frac{S_t}{K})+(r+\frac{\sigma^2}{2})(T-t)\right],
	\nonumber
	\\
	d_2 &= d_1 - \sigma \sqrt{T-t},t\in [0,T).
	\label{eq_call_bsm32}
\end{align}
In (\ref{eq_call_bsm32}), $D_y = D^{(\aleph;\mathcal{C})}$ and $N(x),x \in R$ is the standard normal distribution function.  For a misinformed trader the yield is negative, and thus, in general, if $p^{(\aleph)}_{\Delta t} =\frac{1}{2}+\frac{1}{2} \frac{\lambda^{(\aleph)}}{\sqrt{p_{0}(1-p_{0})}}\sqrt{\Delta t}, \lambda^{(\aleph)}\in R$, the dividend yield $D_y$ in (\ref{eq_call_bsm32}) is given by $D_y = D_y^{(\aleph;\mathcal{C})} = \textup{sign}(\lambda^{(\aleph)})\sigma(\sqrt{\theta^2+4\lambda^{(\aleph)^2}}-\theta)$.

\subsection{Implied information rate $\lambda^{(\aleph)}$}
Here we apply the Black-Sholes option pricing formula to construct the implied trader information intensity surface $\lambda^{(\aleph;implied)} $  and implied probability $p_{\Delta t}^{(\aleph;implied)}=\frac{1}{2} \left( 1+\frac{\lambda^{(\aleph;implied)}}{\sqrt{p_0\left(1-p_0 \right) }}\sqrt{\Delta t}\right)$. We calculate $p_{\Delta t}^{(\aleph;implied)} $  for options with different times to maturity and strike prices.

To this end, we first estimate  $p_0$, as the sample success probability $\hat{p}_0$  of the stock price being “up” for a fixed day. We use one-year of historical log-returns to calculate the sample mean $\hat{\mu}$ as an estimate for $\mu$, and the historical sample standard deviation $\hat{\sigma}$ as an estimate for $\sigma$. Here, we compare the option and spot trader's information of stock returns by applying $\lambda^{(\aleph;implied)}$  and $p_{\Delta t}^{(\aleph;implied)} $. Rather than looking at individual stocks, our analysis will focus on the aggregate stock market. In our analysis, we selected a broad-based market index, the S\&P 500, as measured by the SPDR S\&P 500, which is an exchange-traded fund, as the proxy for the aggregate stock market. We use the 10-year Treasury yield as a proxy for the risk-free rate $r$.  The database includes the period from November 2018 to November 2019. There were 252 observations collected from Yahoo Finance.

We use call option prices on $10/24/2019$ with different expiration dates and strike prices. The expiration date varies from $10/25/2019$ to $01/21/2022$, and the strike price varies from $25$ to $450$ among various call option contracts. The midpoint of the bid and ask is applied in the computation. As the underlying of the call option, the SPY index price was $\$301.6$ on $10/22/2019$. We use the 10-year Treasury yield curve rate on $01/02/2015$ as the risk-free rate, here  $r= 1.801\%$. 
	
\begin{figure}[htb!]
	\centering
	\includegraphics[scale=0.32,angle=0]{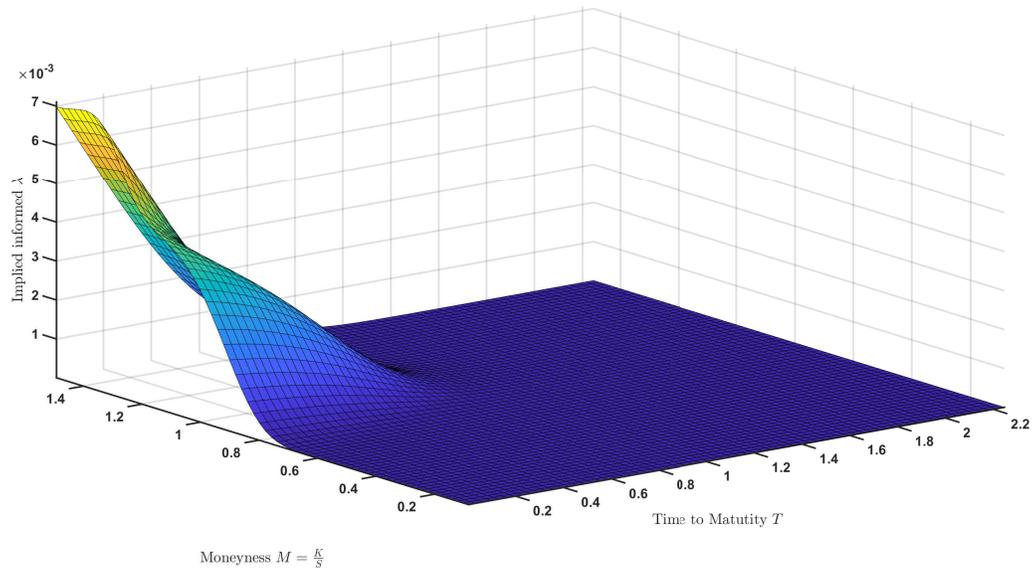}
	\caption{Implied information against time to maturity and moneyness.}
	\label{fig:task2_1}
\end{figure}

Matching the theoretical option prices in (\ref{eq_call_bsm32}) and with the corresponding market prices $C^{(market;\mathcal{C},i)}$ $(S_t,K,T-t)$, $i=1,\ldots,T$, we first construct the implied trader information intensity $\lambda^{(\aleph)}=\lambda^{(\aleph;implied)}$-implied surface. $\lambda^{(\aleph;implied)}$-implied surface is graphed against both a standard measure of moneyness and time to maturity (in years) in Figure \ref{fig:task2_1}. 
	
Figure \ref{fig:task2_1} indicates that at each maturity, the implied trader information intensity of option traders increases as moneyness increases. Where the moneyness varies in $(0, 0.75)$, the surface is flat at point $0$, indicating equal information of spot traders and option traders and efficiency of the markets. Where the moneyness varies in $(0.75, 1.15)$, the value of  $\lambda^{(N;implied)} $ starts increasing from zero to $0.007$. This finding indicates that option traders potentially are more informed when the option is in-the-money. The surface shows that when the option is in-the-money and when it is not a significant out-the-money option, the option traders are more informed.
	
\begin{figure}[htb!]
	\centering
	\includegraphics[scale=0.32,angle=0]{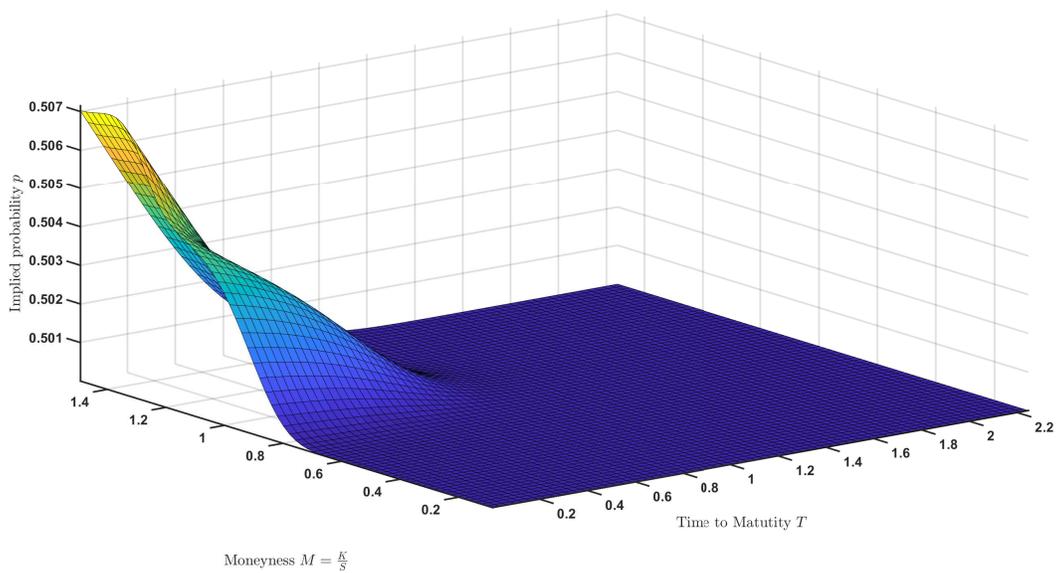}
	\caption{Implied probability against time to maturity and moneyness.}
	\label{fig:task2_2}
\end{figure}

The implied probability $p_{\Delta t}^{(\aleph;implied)} $-surface is graphed against both a standard measure of “moneyness” and time to maturity (in years) in Figure \ref{fig:task2_2}. Recall that values higher than $0.5$ for $p_{\Delta t}^{(\aleph;implied)}$, means that option traders have more information about the mean  $\mu$ of SPY daily return or the option trader is the informed trader. In other words, $p_{\Delta t}^{(\aleph;implied)} >0.5$ means that option traders are potentially informed about the future of the SPY returns. The opposite is true when $p_{\Delta t}^{(\aleph;implied)} \leq 0.5$. 
	
Figure \ref{fig:task2_2} shows that at each maturity, the information of option traders about the mean of SPY daily log-returns increases as moneyness increases. Where the moneyness ranges in (0, 0.75), the surface is flat at point 0.5, indicating that spot traders and option traders are equally informed, and the predictability of the market is equal for both types of traders. Where the moneyness changes in $(0.75, 1.15)$, $p_{\Delta t}^{(\aleph;implied)} $ starts increasing from $0.5$ to $0.507$. Again, this finding indicates that option traders are potentially more informed when the option is in-the-money, a finding consistent with Shirvani (2019).
\section{Option Pricing When The Trader Has Information on The Instantaneous Mean Return}
In this section we assume that trader $\aleph$ has information  about whether the instantaneous mean return of stock $\mathcal{S}$ is above or below the market perceived value $\mu$. First, we notice that this information, is equivalent to $\aleph$'s information about the probability for stock price upturn. Consider the binomial stock price model: $S_{0,n}^{(p_{\Delta t})} = S_0$, and conditionally on $S_{k,n}^{(p_{\Delta t})}$,
\begin{equation}
S_{k+1,n}^{(p_{\Delta t})} = 
	\begin{cases} 
      S_{k+1,n}^{(p_{\Delta t},u)} = S_{k,n}^{(p_{\Delta t})}(1+\mu\Delta t+\sigma\sqrt{\frac{1-p_0}{p_0}}\sqrt{\Delta t}),\; w.p. \; p_0+\delta^{(\aleph)}\sqrt{\Delta t},
      \\
      S_{k+1,n}^{(p_{\Delta t},d)} = S_{k,n}^{(p_{\Delta t})}(1+\mu\Delta t-\sigma\sqrt{\frac{p_0}{1-p_0}}\sqrt{\Delta t}),\; w.p. \; 1-p_0-\delta^{(\aleph)}\sqrt{\Delta t},
      \end{cases}
      \label{eq_updown33}
\end{equation}
$k = 1,...,n,\;n\Delta t = T$, where $\delta^{(\aleph)}\in(\frac{1}{2},1)$, and $o(\Delta t)=0$. Then, $\mathbb{E}\left(\frac{S_{k+1,n}^{(p_{\Delta t})}}{S_{k,n}^{(p_{\Delta t})}}\right) = 1+ \mu^{(\aleph)}\Delta t$, where $\mu^{(\aleph)} = \mu+\frac{\sigma}{\sqrt{p_0(1-p_0)}}\delta^{(\aleph)}$, and $Var\left(\frac{S_{k+1,n}^{(p_{\Delta t})}}{S_{k,n}^{(p_{\Delta t})}}\right) = \sigma^2\Delta t$. Thus, $\aleph$'s belief that the true stock mean return is $\mu^{(\aleph)} = \mu+\frac{\sigma}{\sqrt{p_0(1-p_0)}}\delta^{(\aleph)},\delta^{(\aleph)}\neq 0$, rather than the market perceived value $\mu$, is expressed by $\aleph$'s  belief that the true stock price dynamics is given by: $S_{0,n}^{(p_{\Delta t})} = S_0$, and for $k =0,1,...,n-1,n\Delta t = T$,
\begin{equation}
	S_{k+1,n}^{(p_{\Delta t})} = 
	\begin{cases} 
      S_{k+1,n}^{(p_{\Delta t},u)} = S_{k,n}^{(p_{\Delta t})}(1+\mu^{(\aleph)}\Delta t+\sigma\sqrt{\frac{1-p_0}{p_0}}\sqrt{\Delta t}),\; w.p. \; p_0
      \\
      S_{k+1,n}^{(p_{\Delta t},d)} = S_{k,n}^{(p_{\Delta t})}(1+\mu^{(\aleph)}\Delta t-\sigma\sqrt{\frac{p_0}{1-p_0}}\sqrt{\Delta t}),\; w.p. \; 1-p_0
      \end{cases}
      \label{eq_updown34}
\end{equation}
conditionally on $S_{k,n}^{(p_{\Delta t})}$. $\aleph$ can use pricing model (\ref{eq_updown34}) or, equivalently\footnote{Here “equivalently” means that the binomial option pricing trees in (\ref{eq_updown33}) and (\ref{eq_updown34}) generate the same limiting geometric Brownian motion, as $\Delta t \downarrow 0, n\Delta t = T$. This argument follows from the DPIP.}, can use model (\ref{eq_updown33}). According to (\ref{eq_updown33}), $\aleph$ believes that  the true probability for an upturn is $p_{\Delta t} = p_0 +\delta^{(\aleph)}\sqrt{\Delta t}$ for some $\delta^{(\aleph)} \neq 0$, while the market perceived stock price dynamics is given by (\ref{eq_updown33}) but with $\delta^{(\aleph)} = 0$.

\subsection{Forward contract strategy of a trader with information on instantaneous mean stock return}
We introduce $\aleph$'s strategy of trading forward contracts based on information on the instantaneous mean stock return. If at $k\Delta t$, $\aleph$ believes that the stock mean return is $\mu^{(\aleph)} = \mu+\frac{\sigma}{\sqrt{p_0(1-p_0)}}\delta^{(\aleph)},\delta^{(\aleph)}>0$, he enters $\Delta_{k\Delta t}^{(\aleph, \mu)}$ -- long forwards, with $\Delta_{k\Delta t}^{(\aleph, \mu)} = \frac{N^{(\aleph,\mu)}}{S_{k,n}^{(p_{\Delta t})}},N^{(\aleph,\mu)}>0$, and terminal time $(k+1)\Delta t$. If $\aleph$ believes that the stock mean return is $\mu^{(\aleph)} = \mu+\frac{\sigma}{\sqrt{p_0(1-p_0)}}\delta^{(\aleph)},\delta^{(\aleph)}<0$, he enters $\Delta_{k\Delta t}^{(\aleph, \mu)}$ -- short forwards\footnote{The probability of  $\aleph$'s guess on the stock price direction being correct is assumed to be $p_{\Delta t}^{(\aleph, \mu)}\in(0,1)$.}. In other words, if, at time instance $k \Delta t$, $\aleph$ believes that the true probability for stock price upturn is $p_{\Delta t} = p_0+\delta^{(\aleph)}\sqrt{\Delta t}$ with $\delta^{(\aleph)}>0$, he  bets that the stock price will be $S_{k+1,n}^{(p_{\Delta t},u)} = S_{k,n}^{(p_{\Delta t})}(1+\mu\Delta t+\sigma\sqrt{\frac{1-p_0}{p_0}}\sqrt{\Delta t})$. In this case, $\aleph$ enters $\Delta_{k\Delta t}^{(\aleph, \mu)}$ -- long forward contracts. If $\aleph$ believes that the true probability for the stock price upturn is $p_{\Delta t} = p_0+\delta^{(\aleph)}\sqrt{\Delta t}$ with $\delta^{(\aleph)}<0$, he bets that  the stock price will be $S_{k,n}^{(p_{\Delta t})}(1+\mu\Delta t-\sigma\sqrt{\frac{p_0}{1-p_0}}\sqrt{\Delta t})$. In this case, $\aleph$ enters $\Delta_{k\Delta t}^{(\aleph, \mu)}$ -- short forward contracts. Following the same arguments as in Section 3.1, the conditional mean and variance of $P_{k\Delta t \rightarrow (k+1)\Delta t}^{(\aleph;forward,\mu)}$ are given by
\begin{align}
\mathbb{E}(P_{k\Delta t \rightarrow (k+1)\Delta t}^{(\aleph;forward,\mu)}|S_{k,n}^{(p_{\Delta t})}) 
&= N^{(\aleph,\mu)}(2p_{\Delta t}^{(\aleph,\mu)}-1)\sigma(\theta(2p_{k\Delta t} -1)\Delta t \nonumber
\\
&+ 2 \sqrt{p_{\Delta t}(1-p_{\Delta t})}\sqrt{\Delta t}), \nonumber
\\
Var(P_{k\Delta t \rightarrow (k+1)\Delta t}^{(\aleph;forward,\mu)}|S_{k,n}^{(p_{\Delta t})}) 
&=  N^{(\aleph,\mu)^2}\sigma^2\left(1-4(2p^{(\aleph,\mu)}_{\Delta t}-1)^2p_{\Delta t}(1-p_{\Delta t})\right)\Delta t.
\label{eq_meanvar_36}
\end{align}
By the DPIP, for $P_{k\Delta t \rightarrow (k+1)\Delta t}^{(\aleph;forward,\mu)}$, we must have $\mathbb{E}(P_{k\Delta t \rightarrow (k+1)\Delta t}^{(\aleph;forward,\mu)}) = O(\Delta t)$. Assuming that $\aleph$ is an informed trader, we set $p^{(\aleph,\mu)}_{\Delta t} = \frac{1}{2}(1+\frac{\rho^{(\aleph,\mu)}}{\sqrt{p_{\Delta t}(1-p_{\Delta t})}}\sqrt{\Delta t})$, for some $\rho^{(\aleph,\mu)}>0$. Thus, with $p_{\Delta t} = p_0+\delta^{(\aleph)}\sqrt{\Delta t}$, (\ref{eq_meanvar_36}) is simplified  and has the form:
\begin{align}
\mathbb{E}(P_{k\Delta t \rightarrow (k+1)\Delta t}^{(\aleph;forward,\mu)}) 
&= 2N^{(\aleph,\mu)}\sigma\rho^{(\aleph,\mu)}\Delta t, \nonumber
\\
Var(P_{k\Delta t \rightarrow (k+1)\Delta t}^{(\aleph;forward,\mu)}) 
&=  N^{(\aleph,\mu)^2}\sigma^2\Delta t.
\label{eq_meanvar_37}
\end{align}
In (\ref{eq_meanvar_37}), the mean $\mathbb{E}(P_{k\Delta t \rightarrow (k+1)\Delta t}^{(\aleph;forward,\mu)})$ and the variance $Var(P_{k\Delta t \rightarrow (k+1)\Delta t}^{(\aleph;forward,\mu)})$ do not depend on the actual value of $\delta^{(\aleph)}$ in $\mu^{(\aleph)} = \mu+\frac{\sigma}{\sqrt{p_0(1-p_0)}}\delta^{(\aleph)}$. This is due to the fact that $\aleph$ knows, with probability $p_{\Delta t}^{(\aleph,\mu)}$, whether $\mu^{(\aleph)}$ is above or below $\mu$; that is, $\aleph$ knows $\textup{sign}(\delta^{(\aleph)})$, but not  the value $|\delta^{(\aleph)}|$. 

\subsection{Binomial option pricing when the trader has information on instantaneous mean stock return}
Comparing (\ref{eq_meanvar_22}), (\ref{eq_meanvar_23}) with (\ref{eq_meanvar_36}) and (\ref{eq_meanvar_37}), it becomes clear that $\aleph$'s information (on whether the instantaneous mean stock return is above or below the market perceived value $\mu$) is equivalent to $\aleph$'s information on whether the instantaneous upward probability is above or below the market perceived probability $p_0$. Thus, when $\aleph$ applies the forward strategy, the option pricing formula (\ref{eq_call_bsm32}) is valid with $D_y = D_y^{(\aleph;\mathcal{C},\mu)} = \sigma(\sqrt{\theta^2+4\rho^{(\aleph,\mu)^2}}-\theta)>0$. As the yield $D_y = D_y^{(\aleph;\mathcal{C},\mu)}$ does not depend on $\delta^{(\aleph)}$, the option price (\ref{eq_call_bsm32}) does not depend on $Dev^{(\aleph,\mu)} = \mu^{(\aleph)}-\mu = \frac{\sigma}{p_0(1-p_0)}\delta^{(\aleph)}$. However, consider the case where the binomial option pricing tree is given by: $S_{0,n}^{(\aleph;\mathcal{C},\mu)} = S_0$, and conditionally on $S_{k,n}^{(p_{\Delta t})}$,
\begin{equation}
	S_{k+1,n}^{(\aleph;\mathcal{C},\mu)} = 
	\begin{cases} 
      S_{k+1,n}^{(p_{\Delta t},u)}+N^{(\aleph,\mu)}(S_{k+1,n}^{(p_{\Delta t},u)}-S_{k,n}^{(p_{\Delta t})}e^{r\Delta t}),\; w.p. \; p_{\Delta t}p_{\Delta t}^{(\aleph,\mu)},
      \\
      S_{k+1,n}^{(p_{\Delta t},d)}+N^{(\aleph,\mu)}(S_{k,n}^{(p_{\Delta t})}e^{r\Delta t}-S_{k+1,n}^{(p_{\Delta t},d)}),\; w.p. \; (1-p_{\Delta t})p_{\Delta t}^{(\aleph,\mu)},
      \\
      S_{k+1,n}^{(p_{\Delta t},u)}+N^{(\aleph,\mu)}(S_{k,n}^{(p_{\Delta t})}e^{r\Delta t}-S_{k+1,n}^{(p_{\Delta t},u)}),\; w.p. \; p_{\Delta t}(1-p_{\Delta t}^{(\aleph,\mu)}),
      \\
      S_{k+1,n}^{(p_{\Delta t},d)}+N^{(\aleph,\mu)}(S_{k+1,n}^{(p_{\Delta t},d)}-S_{k,n}^{(p_{\Delta t})}e^{r\Delta t}),\; w.p. \; (1-p_{\Delta t})(1-p_{\Delta t}^{(\aleph,\mu)}),
	\end{cases}
	\label{eq_tree38}
\end{equation}
$k = 0,1,...,n-1,n\Delta t = T$. According to (\ref{eq_op_price27}), the optimal $N^{(\aleph,\mu)}$ is given by $N^{(\aleph,\mu)} = N^{(\aleph,\mu;opt)} = 2\frac{\rho^{(\aleph,\mu)}}{\theta}$, leading to
\begin{align}
	\mathbb{E}(R_{k,n}^{(\aleph;\mathcal{C},\mu)} |S_{k,n}^{(\aleph;\mathcal{C},\mu)}) &= (\mu+4\sigma^2\frac{\rho^{{(\aleph,\mu)}^2}}{\mu-r})\Delta t, \nonumber
	\\
	Var(R_{k,n}^{(\aleph;\mathcal{C},\mu)} |S_{k,n}^{(\aleph;\mathcal{C},\mu)}) &= \sigma^2(1+4\sigma^2\frac{\rho^{{(\aleph,\mu)}^2}}{(\mu-r)^2})\Delta t,
\label{eq_tree39}
\end{align}
where $R_{k,n}^{(\aleph;\mathcal{C},\mu)}=\log(\frac{S_{k+1,n}^{(\aleph;\mathcal{C},\mu)} }{S_{k,n}^{(\aleph;\mathcal{C},\mu)} })$. Consider the binomial option pricing tree: $S_{0,n}^{(p_{\Delta t};\delta,\rho)}=S_0$, and conditionally on $S_{k,n}^{(p_{\Delta t};\delta,\rho)}$,
\begin{equation}
	S_{k+1,n}^{(p_{\Delta t};\delta,\rho)} = 
	\begin{cases} 
      S_{k+1,n}^{(p_{\Delta t},u;\delta,\rho)} = S_{k,n}^{(p_{\Delta t};\delta,\rho)}(1+v^{(\aleph)}_{\Delta t}\Delta t+S^{(\aleph)}_{\Delta t} \sqrt{\frac{1-p_{\Delta t}}{p_{\Delta t}}}\sqrt{\Delta t}),\; w.p. \; p_{\Delta t},
      \\
      S_{k+1,n}^{(p_{\Delta t},d;\delta,\rho)} = S_{k,n}^{(p_{\Delta t};\delta,\rho)}(1+v^{(\aleph)}_{\Delta t}\Delta -S^{(\aleph)}_{\Delta t} \sqrt{\frac{p_{\Delta t}}{1-p_{\Delta t}}}\sqrt{\Delta t}),\;\;\; w.p. \; 1-p_{\Delta t},
      \end{cases}
      \label{eq_tree40}
\end{equation}
for $k = 0,1,...,n-1$. In (\ref{eq_tree40}), $v^{(\aleph)}_{\Delta t} = \mu+4\sigma^2\frac{\rho^{{(\aleph,\mu)}^2}}{\mu-r},S^{(\aleph)}_{\Delta t} = \sigma\sqrt{1+4\sigma^2\frac{\rho^{{(\aleph,\mu)}^2}}{(\mu-r)^2}}$. By the DPIP,  the trees (\ref{eq_tree38}) and (\ref{eq_tree40}) have the same limiting pricing process as $\Delta t \rightarrow 0$. The limiting process is independent of $p_{\Delta t} = p_0+\delta^{(\aleph)}\sqrt{\Delta t}\in(0,1)$, and thus the information about $\delta^{(\aleph)}$ and $\mu^{(\aleph)} = \mu+\frac{\sigma}{\sqrt{p_0(1-p_0)}}\delta^{(\aleph)}$ will be lost. However, for a fixed trading frequency $\Delta t$, the risk--neutral probabilities $q_{\Delta t}^{(\aleph;\delta,\rho)}$ and $1-q_{\Delta t}^{(\aleph;\delta,\rho)}$ corresponding to the tree (\ref{eq_tree40}) are given by (\ref{eq_tree_q_16}):
\begin{equation}
	q_{\Delta t}^{(\aleph;\delta,\rho)} = p_{\Delta t}- \theta^{(\aleph;\delta,\rho)}_{\Delta t} \sqrt{p_{\Delta t}(1-p_{\Delta t})}\sqrt{\Delta t},
	\label{eq_q_41}
\end{equation}
where $\theta^{(\aleph;\delta,\rho)}_{\Delta t} = \sqrt{\theta^2+4\rho^{(\aleph,\mu)^2}}$. Thus, the binomial option price process
\begin{equation}
	f_{k,n} = e^{-r \Delta t}(q_{\Delta t}^{(\aleph;\delta,\rho)} f_{k+1,n}^{(u)}+(1-q_{\Delta t}^{(\aleph;\delta,\rho)})f_{k+1,n}^{(d)}), k = 0,...,n
	\label{eq_tree42}
\end{equation}
depends on $\delta^{(\aleph)}$, and thus on $Dev^{(\aleph,\mu)} = \frac{\sigma}{\sqrt{p_0(1-p_0)}}\delta^{(\aleph)}$ as well.

\subsection{Implied $Dev^{(\aleph,\mu)}$}
Similar to Section 2.4, we introduce the concept of \textit{implied $Dev^{(\aleph,\mu)}$-surface} using the following numerical example. 

Again, our dataset is collected from daily closing prices for the SPY and  the call option contracts XSP, and we use the 10-year Treasury yield as an approximation of the riskless rate $r$. By setting $5/15/2020$ as the starting date, we get $S_0 =\$ 286.28$ and $r = 0.64\%$. With one-year back trading data, we get $\hat{p}_0 = 0.56$, $\hat{\mu}_0 = 1.80\times 10^{-4}$, and $\hat{\sigma}_0 = 0.02$.

In the previous section we showed that $Dev^{(\aleph,\mu)} = \frac{\sigma}{\sqrt{p_0(1-p_0)}}\delta^{(\aleph)}$. After initiating parameters, the task becomes finding the implied $\delta^{(\aleph)}$-surface. According to (\ref{eq_tree38}), (\ref{eq_tree40}), (\ref{eq_q_41}), and (\ref{eq_tree42}), we build up a binomial option pricing tree involving unknown parameters $\delta^{(\aleph)}>0$ and $\rho^{(\aleph,\mu)}>0$. To construct the implied $\delta^{(\aleph)}$-surface, we want to fix $\rho^{(\aleph,\mu)}$. With $\delta^{(\aleph)}\in(\frac{1}{2},1)$, we set $\delta_j^{(\aleph)} = 0.50,0.51,...,1$, then we find the optimal $\rho^{(\aleph,\mu;opt)} = $
\begin{align*}
	\textup{arg\,min} \sum_{i=1}^{I} \sum_{j=1}^{J} \left(\frac{C^{(\aleph;\mathcal{C},i)}(S_0,K,T,r,\hat{p},\hat{\sigma},\delta_j^{(\aleph)},\rho^{(\aleph,\mu)})-C^{(market;\mathcal{C},i)}(S_0,K,T,r,\hat{p},\hat{\sigma})}{C^{(market;\mathcal{C},i)}(S_0,K,T,r,\hat{p},\hat{\sigma})}\right)^2.
\end{align*}
In this numerical example, we set $\rho^{(\aleph,\mu;opt)} = 0.49$. Then, we calculate the implied $\delta^{(\aleph)}$-surface and implied $Dev^{(\aleph)}$-surface. 

\begin{figure}[htb!]
	\centering
	\includegraphics[scale=0.32,angle=0]{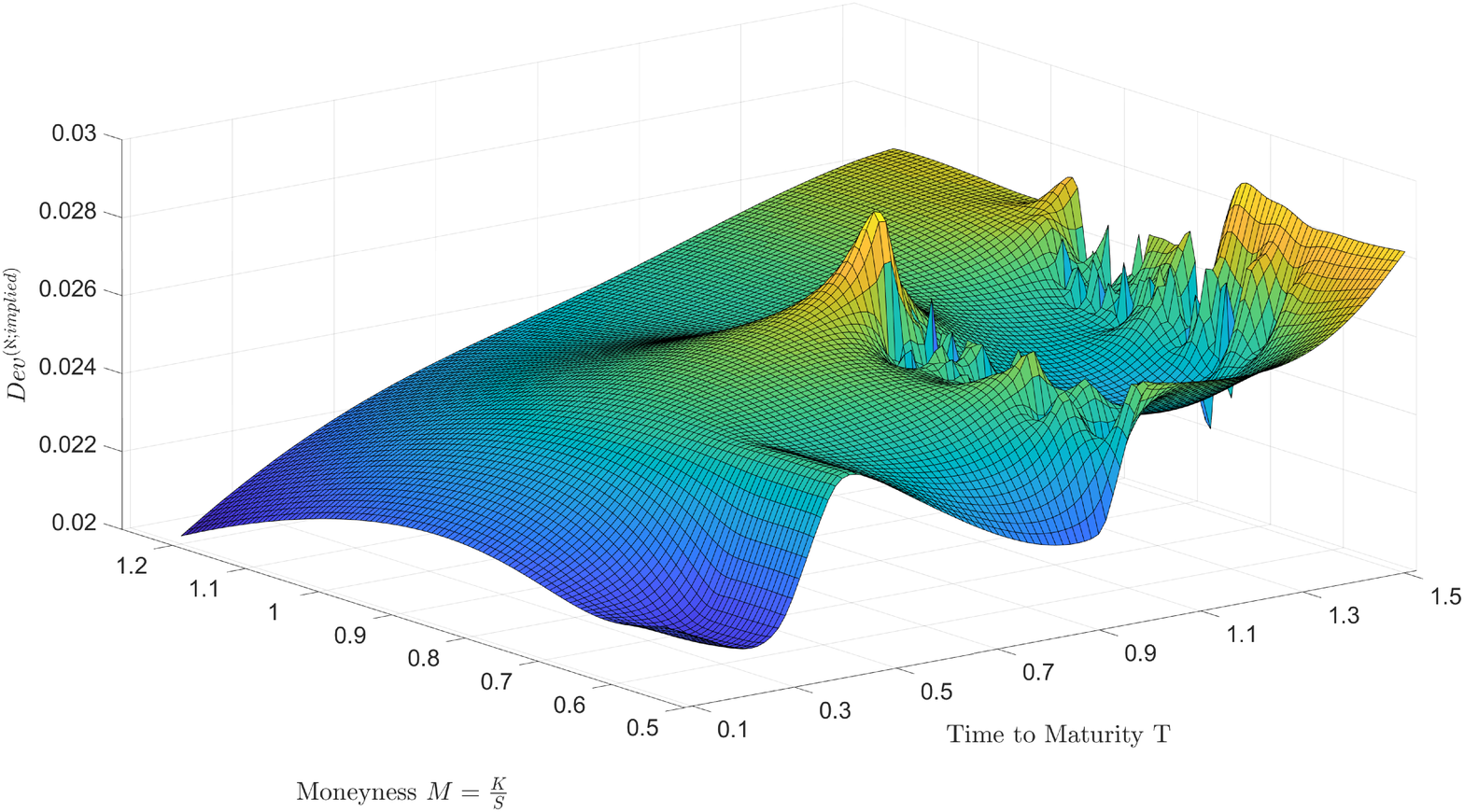}
	\caption{Implied $Dev^{(\aleph,\mu)}$-surface against time to maturity and moneyness.}
	\label{fig:task3_1}
\end{figure}

FigureFigure \ref{fig:task3_1} shows the implied $Dev^{(\aleph)}$-surface against “moneyness”  and time to maturity $T$ in years. Here, $M \in [0.5,1.2]$, $T\in[0.1,1.5]$, and $Dev^{(\aleph;implied)}\in [0.0202,0.0289]$. We observe that the value of $Dev^{(\aleph;implied)}$ fluctuates between $0.02$ and $0.03$ on different call option contracts for SPY in this numerical example. However, for fixed $M$, we can still capture the trend of increments of $Dev^{(\aleph;implied)}$ as $T$ increases. Recall the definition of implied $Dev^{(\aleph)}$: $Dev^{(\aleph,\mu)} = \mu^{(\aleph)}-\mu = \frac{\sigma}{p_0(1-p_0)}\delta^{(\aleph)}$. This fact indicates that option traders believe the SPY will go “up” in the future.
\section{Option Pricing When The Trader Has Information on The Underlying Asset with Price Process Following Continuous Diffusion}
If $\aleph$ has information whether the true volatility of stock $\mathcal{S}$ is above or below the market perceived value $\sigma$, the trader should trade the $\mathcal{S}$-volatility\footnote{If $\mathcal{S}$ is SPDR (S$\&$P ETF Trust, SPY, State Street Global Advisors), then VIXY (VIX Short-Term Futures ETF)-tracks S$\&$P500 volatility, traded as S$\&$P 500 VIX Short-Term Futures Index. Volatility indices on stock indices, individual equities, currencies and interest rates are traded at the CBOE, \url{http://www.cboe.com/products/vix-index-volatility/volatility-indexes.}}. If $\aleph$ has information whether the true interest rate is above or below the market perceived value $r$, the trader should invest in a money market ETF. As the volatility and interest rate dynamics are generally mean reverting, we next extend our option pricing model for informed traders in financial markets driven by a continuous-diffusion process.

\subsection{KSRF-binomial pricing tree with time-varying parameters}
We start with the KSRF-binomial model for the continuous-diffusion price process. Consider the continuous-diffusion market $(\mathcal{S},\mathcal{B},\mathcal{C})$ within the BSM framework with stock $\mathcal{S}$, bond $\mathcal{B}$, and ECC $\mathcal{C}$. The stock price dynamics follows a continuous-diffusion process $\mathbb{S} = \{S_t,t\in [0,T]\}$, where
\begin{equation}
	S_t =  S_0\exp\left\{\int_{0}^{t}(\mu_u-\frac{1}{2}\sigma_u^2)du+\int_{0}^{t}\sigma_u dB_u\right\},t\in [0,T] ,S_0>0
	\label{eq_op_price_43}
\end{equation}
defined on $(\Omega, \mathbb{F} = \{\mathcal{F}_t\}_{t\in [0,T]},\mathbb{P})$ with filtration $\mathbb{F}$, generated by the BM $B_t,t\in [0,T]$. The instantaneous mean function $\mu_t>0, t\in [0,T]$ and the volatility function $\sigma_t>0,t\in [0,T]$ are  deterministic and have continuous derivatives on $[0,T]$. The bond price is given by
\begin{equation}
	\beta_t =\beta_0\exp\left\{\int_{0}^{t}r_u du\right\},t\in[0,T],\beta_0>0,
	\label{eq_op_price_44}
\end{equation}
where the instantaneous riskless rate $r_t,t\in[0,T]$, has continuous derivative on $[0,T]$, and $0<r_t<\mu_t,t\in[0,T]$. The ECC with underlying asset $\mathcal{S}$ has terminal (expiration) time $T>0$, and terminal payoff $g(S_T)$. Let $n \Delta t= T,n \in N=\{1,2,…\}$, and $\epsilon_{k\Delta t}^{(p)},k=1,…,n$ be a sequence of independent Bernoulli random variables with $\mathbb{P}(\epsilon_{(k+1)\Delta t}^{(p)}=1)=1-\mathbb{P}(\epsilon_{(k+1)\Delta t}^{(p)}=0)=p_{k\Delta t},k=0,1,…,n-1$,  where $p_t\in(0,1),t\in[0,T]$ has continuous  first derivative. Consider the KSRF-binomial price dynamics :
\begin{equation}
	S_{(k+1)\Delta t,n}^{(p)} = 
	\begin{cases} 
      S_{(k+1)\Delta t,n}^{(p,u)} = S_{k\Delta t,n}^{(p)}(1+\mu_{k\Delta t}\Delta t+\sigma_{k\Delta t} \sqrt{\frac{1-p_{k\Delta t}}{p_{k\Delta t}}}\sqrt{\Delta t}),\; \textrm{if} \; \epsilon_{k\Delta t}^{(p)} = 1,
      \\
      S_{(k+1)\Delta t,n}^{(p,d)} = S_{k\Delta t,n}^{(p)}(1+\mu_{k\Delta t}\Delta -\sigma_{k\Delta t} \sqrt{\frac{p_{k\Delta t}}{1-p_{k\Delta t}}}\sqrt{\Delta t}),\;\; \textrm{if} \; \epsilon_{k\Delta t}^{(p)} = 0,
      \end{cases}
      \label{eq_updown44}
\end{equation}
for $k = 0,1,...,n-1$ and $S_{0,n}^{(p)} = S_0$. With $o(\Delta t) = 0$, (\ref{eq_updown44}) is a recombined tree, and $\mathbb{E}[(\frac{S_{(k+1)\Delta t,n}^{(p)}}{S_{k\Delta t,n}^{(p)}})^{\gamma}|{S_{k\Delta t,n}^{(p)}}] = \mathbb{E}[(\frac{S_{(k+1)\Delta t}}{S_{k\Delta t}})^{\gamma}|{S_{k\Delta t}}] = 1+\gamma (\mu_{k\Delta t}+\frac{\gamma-1}{2}\sigma^2_{k\Delta t})$ for all $\gamma>0$. Set $\mathbb{S}^{(n,p)} = \{S_t^{(n,p)},t\in[0,T]\}$, where $S_t^{(n,p)} = S_{k\Delta t,n}^{(p)}$ for $t \in [k\Delta t, (k+1)\Delta t),k = 0,1,...,n-1, S_T^{(n,p)} = S_{n,n}^{(p)}$. Then, as $n \uparrow \infty$, $\mathbb{S}^{(n,p)}$ weakly converges in $\mathcal{D}[0,T]$ to $\mathbb{S}$.\footnote{See Proposition 3 in Davydov \& Rotar (2008) and Kim et al. (2019).} The risk-neutral probabilities $q_{k\Delta t},k = 0,1,...,n-1$, are $q_{k\Delta t} = p_{k\Delta t}-\theta_{k\Delta t} \sqrt{p_{k\Delta t}(1-p_{k\Delta t})\Delta t}$, with $o(\Delta t) = 0$ and $\theta_t = \frac{\mu_t-r_t}{\sigma_t},t \in [0,T]$.

The risk-neutral tree pricing tree is given by
\begin{equation}
	S_{(k+1)\Delta t,n}^{(q)}= 
	\begin{cases}
	S_{(k+1)\Delta t,n}^{(q,u)}  = S_{k\Delta t,n}^{(q)} (1+r_{k\Delta t}\Delta t+\sigma_{k\Delta t}\sqrt{\frac{1-q_{k\Delta t}}{q_{k\Delta t}}}\sqrt{\Delta t}),\textrm{if} \; \epsilon_{(k+1)\Delta t}^{(q)} = 1,
	\\
	S_{(k+1)\Delta t,n}^{(q,d)}  = S_{k\Delta t,n}^{(q)} (1+r_{k\Delta t}\Delta t-\sigma_{k\Delta t}\sqrt{\frac{q_{k\Delta t}}{1-q_{k\Delta t}}}\sqrt{\Delta t}),\textrm{if} \; \epsilon_{(k+1)\Delta t}^{(q)} = 0,
	\end{cases}
	\label{eq_option_q45}
\end{equation}
for $k = 0,1,...,n-1$ and $S_{0,n}^{(q)} = S_0$. In (\ref{eq_option_q45}), $\epsilon_{k\Delta t}^{(q)},k=1,...,n$ is a sequence of independent Bernoulli random variables with $\mathbb{P}(\epsilon_{(k+1)\Delta t}^{(q)}=1)=1-\mathbb{P}(\epsilon_{(k+1)\Delta t}^{(q)}=0)=q_{k\Delta t},k=0,1,…,n-1$. Set $\mathbb{S}^{(n,q)} = \{S_t^{(n,q)},t\in[0,T]\}$, where $S_t^{(n,q)}=S_{k,n}^{(q)}$ for $t \in [k\Delta t, (k+1)\Delta t),k = 0,1,...,n-1,S^{(n,q)}_T = S^{(q)}_{n,n}$. Then, as $n \uparrow \infty$, $\mathbb{S}^{(n,q)}$ weakly converges in $\mathcal{D}[0,T]$ to $\mathbb{S}^{(\mathbb{Q})} = \{S_t^{(\mathbb{Q})},t\in [0,T]\}$, where $S_t^{(\mathbb{Q})} = S_0\exp\{\int_{0}^{t}(r_u-\frac{1}{2}\sigma_u^2)du+\int_{0}^{t}\sigma_u dB_u^{\mathbb{Q}}\},t\in [0,T]$, where $B_u^{\mathbb{Q}}$ is a BM on $(\Omega,\mathbb{F} = \{\mathcal{F}_t\}_{t \geq 0}, \mathbb{Q})$, and $\mathbb{Q}$ is the unique equivalent martingale measure.\footnote{See Chapter 6 in  Duffie (2001).}

\subsection{Forward contract strategy for a trader with information on stock price direction in the KSRF-pricing tree with time-varying parameters}
At any time $k\Delta t,k=0,...,n,n\Delta t = T$, $\aleph$ makes independent bets, which are modeled as independent Bernoulli trials $\eta^{(\aleph)}_{(k+1)\Delta t,n},k=0,...,n-1,\mathbb{P}(\eta^{(\aleph)}_{(k+1)\Delta t,n} = 1) = 1-\mathbb{P}(\eta^{(\aleph)}_{(k+1)\Delta t,n} = 0) = p^{\aleph}_{k\Delta t} \in (\frac{1}{2},1)$. The function $p^{\aleph}_t\in (\frac{1}{2},1),t\in[0,T]$ is assumed to have continuous first derivative on $[0,T]$. If at $k \Delta t$, $\aleph$ believes that $\epsilon^{(p)}_{(k+1)\Delta t} = 1$ will happen, he takes a long position in $\Delta _{k\Delta t}^{(\aleph,p)} = \frac{N _{k\Delta t}^{(\aleph,p)}}{S _{k\Delta t,n}^{(p)}}$-- forward contracts for some $N_{k\Delta t}^{(\aleph,p)}>0$. The function $N_{t}^{(\aleph,p)}>0,t\in[0,T]$ is assumed to have continuous first derivative on $[0,T]$. The maturity of the forwards is $(k+1)\Delta t$. If at $k\Delta t$, $\aleph$ believes that $\epsilon^{(p)}_{(k+1)\Delta t} = 0$ will happen, he takes a short position in $\Delta_{k\Delta t}^{(\aleph,p)}$-- forward contracts at maturity $(k+1)\Delta t$. The overall payoff of $\aleph$'s forward contract positions is given by
\begin{equation} 
P_{k\Delta t \rightarrow (k+1)\Delta t}^{(\aleph,p;forward)}= \Delta _{k\Delta t}^{(\aleph,p)}
\begin{cases}
	  (S_{(k+1)\Delta t,n}^{(p,u)}-S_{k\Delta t,n}^{(p)}e^{r_{\Delta t}\Delta t}),\textrm{if} \; \epsilon_{(k+1)\Delta t}^{(p)} = 1, \eta_{(k+1)\Delta t,n}^{(\aleph)} = 1, \\
      (S_{k\Delta t,n}^{(p)}e^{r_{\Delta t}\Delta t}-S_{(k+1)\Delta t,n}^{(p,d)}),\textrm{if} \; \epsilon_{(k+1)\Delta t}^{(p)} = 0, \eta_{(k+1)\Delta t,n}^{(\aleph)} = 1,
      \\
      (S_{k\Delta t,n}^{(p)}e^{r_{\Delta t}\Delta t}-S_{(k+1)\Delta t,n}^{(p,u)}),\textrm{if} \; \epsilon_{(k+1)\Delta t}^{(p)} = 1, \eta_{(k+1)\Delta t,n}^{(\aleph)} = 0,
      \\
      (S_{(k+1)\Delta t,n}^{(p,d)}-S_{k\Delta t,n}^{(p)}e^{r_{\Delta t}\Delta t}),\textrm{if} \; \epsilon_{(k+1)\Delta t}^{(p)} = 0, \eta_{(k+1)\Delta t,n}^{(\aleph)} = 0.
\end{cases}
\end{equation}
The conditional mean and variance of $P_{k\Delta t \rightarrow (k+1)\Delta t}^{(\aleph,p;forward)}$ are given by
\begin{align*}
\mathbb{E}(P_{k\Delta t \rightarrow (k+1)\Delta t}^{(\aleph,p;forward)}|S^{(p)}_{k\Delta t,n}) &= N_{k\Delta t}^{(\aleph,p)}(2p_{k\Delta t}^{\aleph}-1)\sigma_{k\Delta t}(\theta(2p_{k\Delta t} -1)\Delta t \nonumber
\\
&+ 2 \sqrt{p_{k\Delta t}(1-p_{k\Delta t})}\sqrt{\Delta t}), \nonumber
\\
Var(P_{k\Delta t \rightarrow (k+1)\Delta t}^{(\aleph,p;forward)}|S^{(p)}_{k\Delta t,n})&=  N_{k\Delta t}^{(\aleph,p)^2}\sigma_{k\Delta t}^2\left(1-4(2p^{\aleph}_{k\Delta t}-1)^2p_{k\Delta t}(1-p_{k\Delta t})\right)\Delta t.
\end{align*}
By the DPIP\footnote{See Davydov \& Rotar (2008).}, we should have $\mathbb{E}(P_{k\Delta t \rightarrow (k+1)\Delta t}^{(\aleph,p;forward)}|S^{(p)}_{k\Delta t,n}) = O(\Delta t)$, and $Var(P_{k\Delta t \rightarrow (k+1)\Delta t}^{(\aleph,p;forward)}|S^{(p)}_{k\Delta t,n}) = O(\Delta t)$. To guarantee that, we set $p^{\aleph}_{k\Delta t} = \frac{1}{2}(1+\psi^{(\aleph)}_{k\Delta t}\sqrt{\Delta t})$, for some $\psi^{(\aleph)}_{t}>0,t\in[0,T]$.\footnote{The case of a misinformed trader can be considered in a similar manner. A misinformed trader with $\psi^{(\aleph)}<0$,  trades long-forward  (resp. short-forward) when the informed trader with $(-\psi^{(\aleph)})>0$, trades short-forward (resp. long-forward).  A noisy trader will not trade any forward contracts, as he has no information about stock price direction.} It is assumed that $\psi^{(\aleph)}_t,t\in[0,T]$ has continuous first derivative on $[0,T]$. With $p^{\aleph}_{k\Delta t} = \frac{1}{2}(1+\psi^{(\aleph)}_{k\Delta t}\sqrt{\Delta t})$, and $o(\Delta t)=0$, we have
\begin{align*}
\mathbb{E}(P_{k\Delta t \rightarrow (k+1)\Delta t}^{(\aleph,p;forward)}|S^{(p)}_{k\Delta t,n}) &= 2N_{k\Delta t}^{(\aleph,p)}\psi_{k\Delta t}^{(\aleph)}\sigma_{k\Delta t}\sqrt{p_{k\Delta t}(1-p_{k\Delta t})}\Delta t, \nonumber
\\
Var(P_{k\Delta t \rightarrow (k+1)\Delta t}^{(\aleph,p;forward)}|S^{(p)}_{k\Delta t,n})&=  N_{k\Delta t}^{(\aleph,p)^2}\sigma_{k\Delta t}^2\Delta t.
\end{align*}
The \textit{instantaneous information ratio} is given by
\begin{align*}
	IR(P_{k\Delta t \rightarrow (k+1)\Delta t}^{(\aleph,p;forward)}|S^{(p)}_{k\Delta t,n})  &= \frac{\mathbb{E}(P_{k\Delta t \rightarrow (k+1)\Delta t}^{(\aleph,p;forward)}|S^{(p)}_{k\Delta t,n})}{\sqrt{\Delta t}\sqrt{Var(P_{k\Delta t \rightarrow (k+1)\Delta t}^{(\aleph,p;forward)}|S^{(p)}_{k\Delta t,n})}} 
	\\
	&= 2\psi_{k\Delta t}^{(\aleph)}\sqrt{p_{k\Delta t}(1-p_{k\Delta t})}. 
\end{align*}

\subsection{Option pricing for a trader with information on the stock price direction in the KSRF-pricing tree with time-varying parameters}
Suppose now that $\aleph$ is taking a short position in the option contract in the BSM market $(\mathcal{S},\mathcal{B},\mathcal{C})$. The stock price dynamics $S_t = S_t^{(\mu,\sigma)},t \geq 0$, are given by (\ref{eq_op_price_43}), the bond price $\beta_t, t \geq 0$ is given by (\ref{eq_op_price_44}), and the call option $\mathcal{C}$ has price $f_t = f(S_t,t),t \in [0,T]$ with terminal payoff $f_T = g(S_T)$. When $\aleph$ trades the stock $\mathcal{S}$, hedging his short position in $\mathcal{C}$, $\aleph$ simultaneously executes his forward strategy. $\aleph$'s trading strategy (a combination of forward trading with trading the stock)  leads to an enhanced price process, the dynamics of which can be expressed as follows:
\begin{equation}
	S_{k+1,n}^{(\aleph,p;\mathcal{C})}=\begin{cases} 
      S_{(k+1)\Delta t,n}^{(p,u)}+N_{k\Delta t}^{(\aleph,p)}(S_{(k+1)\Delta t,n}^{(p,u)}-S_{k\Delta t,n}^{(p)}e^{r_{k\Delta t}\Delta t}), \textrm{if} \; \zeta_{k+1,n}^{(p_{\Delta t})} = 1, \eta_{k+1,n}^{(\aleph)} = 1, \\
      S_{(k+1)\Delta t,n}^{(p,d)}+N_{k\Delta t}^{(\aleph,p)}(S_{k\Delta t,n}^{(p)}e^{r_{k\Delta t}\Delta t}-S_{(k+1)\Delta t,n}^{(p,d)}), \textrm{if} \; \zeta_{k+1,n}^{(p_{\Delta t})} = 0, \eta_{k+1,n}^{(\aleph)} = 1, \\
      S_{(k+1)\Delta t,n}^{(p,u)}+N_{k\Delta t}^{(\aleph,p)}(S_{k\Delta t,n}^{(p)}e^{r_{k\Delta t}\Delta t}-S_{(k+1)\Delta t,n}^{(p,u)}), \textrm{if} \; \zeta_{k+1,n}^{(p_{\Delta t})} = 1, \eta_{k+1,n}^{(\aleph)} = 0, \\
      S_{(k+1)\Delta t,n}^{(p,d)}+N_{k\Delta t}^{(\aleph,p)}(S_{(k+1)\Delta t,n}^{(p,d)}-S_{k\Delta t,n}^{(p)}e^{r_{k\Delta t}\Delta t}), \textrm{if} \; \zeta_{k+1,n}^{(p_{\Delta t})} = 0, \eta_{k+1,n}^{(\aleph)} = 0,
	\end{cases}
\end{equation}
$k=0,1,...,n-1,n\Delta t = T$.\footnote{With every single share of traded stock with price  $S_{k,n}^{(p_{\Delta t})}$  at $k \Delta t$, $\aleph$ simultaneously enters $N^{(\aleph)}$-- forward contracts. The forward contracts are long or short, depending on $\aleph$'s view on stock price direction in time period $[k\Delta t, (k+1)\Delta t]$.} At time $k\Delta t$, it costs nothing to enter a forward contract with terminal time $(k+1)\Delta t$. Thus, setting $R_{k,n}^{(\aleph,p;\mathcal{C})}=\log(\frac{S_{k+1,n}^{(\aleph,p;\mathcal{C})} }{S_{k,n}^{(\aleph,p;\mathcal{C})} })$, it follows that
\begin{align*}
	\mathbb{E}(R_{k,n}^{(\aleph,p;\mathcal{C})}|S_{k,n}^{(\aleph,p;\mathcal{C})}) &= 
	(\mu_{k\Delta t}+N_{k\Delta t}^{(\aleph,p)}(\mu_{k\Delta t}-r_{k\Delta t})(2p_{k\Delta t}-1)(2p_{k\Delta t}^{\aleph}-1))\Delta t
	\\
	&+ 2N_{k\Delta t}^{(\aleph,p)}\sigma_{k\Delta t}\sqrt{p_{k\Delta t}(1-p_{k\Delta t})}(2p_{k\Delta t}^{\aleph}-1)\sqrt{\Delta t},
	\\
	Var(R_{k,n}^{(\aleph,p;\mathcal{C})}|S_{k,n}^{(\aleph,p;\mathcal{C})}) &= \sigma^2_{k\Delta t}((1+N_{k\Delta t}^{(\aleph,p)^2})-2N_{k\Delta t}^{(\aleph,p)}(2p_{k\Delta t}-1)(2p_{k\Delta t}^{\aleph}-1))\Delta t
	\\
	&-\sigma^2_{k\Delta t}4N_{k\Delta t}^{(\aleph,p)^2}p_{k\Delta t}(1-p_{k\Delta t})(2p_{k\Delta t}^{\aleph}-1)^2\Delta t.
\end{align*}

As discussed in Section 5.2, we set $p^{\aleph}_{k\Delta t} = \frac{1}{2}(1+\psi^{(\aleph)}_{k\Delta t}\sqrt{\Delta t})$. Then with $o(\Delta t) = 0$, $\mathbb{E}(R_{k,n}^{(\aleph,p;\mathcal{C})}|S_{k,n}^{(\aleph,p;\mathcal{C})}) = (\mu_{k\Delta t}+2N_{k\Delta t}^{(\aleph,p)}\sigma_{k\Delta t}\sqrt{p_{k\Delta t}(1-p_{k\Delta t})}\psi_{k\Delta t}^{(\aleph)})\Delta t$, and $Var(R_{k,n}^{(\aleph,p;\mathcal{C})}|S_{k,n}^{(\aleph,p;\mathcal{C})}) = \sigma^2_{k\Delta t}(1+N_{k\Delta t}^{(\aleph,p)^2})\Delta t$. The instantaneous market price of risk is given by 
\begin{align*}
	\Theta(R_{k,n}^{(\aleph,p;\mathcal{C})} |S_{k,n}^{(\aleph,p;\mathcal{C})} ) = \frac{\theta_{k\Delta t}+2N_{k\Delta t}^{(\aleph,p)}\sqrt{p_{k\Delta t}(1-p_{k\Delta t})} \psi_{k\Delta t}^{(\aleph)}}{\sqrt{1+N_{k\Delta t}^{(\aleph,p)^2}}}.
\end{align*}
Then, the optimal $N_{k\Delta t}^{(\aleph,p)}$ maximizing $\Theta(R_{k,n}^{(\aleph,p;\mathcal{C})} |S_{k,n}^{(\aleph,p;\mathcal{C})} ) $, is $N_{k\Delta t}^{(\aleph,p)} = N_{k\Delta t}^{(\aleph,p;opt)} = 2\frac{\psi_{k\Delta t}^{(\aleph)}}{\theta_{k\Delta t}}\sqrt{p_{k\Delta t}(1-p_{k\Delta t})}$,\footnote{By assumption, $\mu_t>r_t>0$, and thus, $\theta_t = \frac{\mu_t-r_t}{\sigma_t}>0,t\in[0,T]$.} and the optimal instantaneous market price of risk is 
\begin{align*}
	\Theta(R_{k,n}^{(\aleph,p;\mathcal{C})} |S_{k,n}^{(\aleph,p;\mathcal{C})} )  = \Theta^{(opt)}(R_{k,n}^{(\aleph,p;\mathcal{C})} |S_{k,n}^{(\aleph,p;\mathcal{C})} ) = \sqrt{\theta^2+4p_{k\Delta t}(1-p_{k\Delta t})\psi_{k\Delta t}^{(\aleph)^2}}.
	\end{align*} 
With $N_{k\Delta t}^{(\aleph,p)} = N_{k\Delta t}^{(\aleph,p;opt)}$, $\mathbb{E}(R_{k,n}^{(\aleph,p;\mathcal{C})}|S_{k,n}^{(\aleph;\mathcal{C})}) 
= (\mu_{k\Delta t}+4\sigma_{k\Delta t}p_{k\Delta t}(1-p_{k\Delta t})\frac{\psi_{k\Delta t}^{(\aleph)^2}}{\theta_{k\Delta t}})\Delta t$, and $Var(R_{k,n}^{(\aleph,p;\mathcal{C})}|S_{k,n}^{(\aleph;\mathcal{C})}) 
= \sigma^2_{k\Delta t}(1+4p_{k\Delta t}(1-p_{k\Delta t})\frac{\psi_{k\Delta t}^{(\aleph)^2}}{\theta^2_{k\Delta t}})\Delta t$.

Next, we consider the limiting behavior of $\{S_{k,n}^{(\aleph,p;\mathcal{C})},k = 0,...,n\}$ as $\Delta t \downarrow 0$. By the DPIP, it follows that, in the limit $\Delta t \downarrow 0$, $\aleph$ hedges his short derivative position using the price process $S_t^{(\aleph,p;\mathcal{C})},t\geq 0$,
\begin{align}
	S_t^{(\aleph,p;\mathcal{C})} &=  S_0\exp\{(\mu_t^{(\aleph,p;\mathcal{C})}-\frac{1}{2}\sigma_t^{(\aleph,p;\mathcal{C})^2})t+\sigma_t^{(\aleph,p;\mathcal{C})} B_t\},
\end{align}
where $\mu_t^{(\aleph,p;\mathcal{C})} = \mu_t+4\sigma_t p_t(1-p_t)\frac{\psi_t^{(\aleph)^2}}{\theta_t},\; \sigma_t^{(\aleph,p;\mathcal{C})} = \sigma_t\sqrt{1+4p_t(1-p_t)\frac{\psi_t^{(\aleph)^2}}{\theta_t^2}}$.
$\aleph$ forms his instantaneous riskless replicating portfolio $\Pi^{(\aleph,p;\mathcal{C})}_t =a_t^{(\aleph,p;\mathcal{C})} S_t^{(\aleph,p;\mathcal{C})}+b_t^{(\aleph,p;\mathcal{C})} \beta_t = f_t = f(S_t^{(\aleph,p;\mathcal{C})},t),t\in[0,T)$ with $df_t = d\Pi^{(\aleph,p;\mathcal{C})}_t =a_t^{(\aleph,p;\mathcal{C})} dS_t^{(\aleph,p;\mathcal{C})}+b_t^{(\aleph,p;\mathcal{C})} d\beta_t$. As in Section 3.2, $f(x,t),x>0,t\in[0,T)$, satisfies the PDE
\begin{align}
\frac{\partial f(x,t)}{\partial t}+(r_t-D_{y,t}^{(\aleph,p;\mathcal{C})})x \frac{\partial f(x,t)}{\partial x}+\frac{1}{2}\sigma_t^2 x^2 \frac{\partial^2 f(x,t)}{\partial x^2}-r_tf(x,t)= 0, 
\label{eq_partial_49}
\end{align}
where $x>0,t\in[0,T)$. And $f(x,T) = g(x)$ is the boundary condition . The dividend yield $D_{y,t}^{(\aleph,p;\mathcal{C})}$ in (\ref{eq_partial_49}) is given by $D_{y,t}^{(\aleph,p;\mathcal{C})} = \sigma_t(\theta_t^{(\aleph,p;\mathcal{C})}-\theta_t)$, where
\begin{align*}
\theta_t^{(\aleph,p;\mathcal{C})} = \frac{\mu_t^{(\aleph,p;\mathcal{C})}-r_t}{\sigma_t^{(\aleph,p;\mathcal{C})}} = \theta_t+\left(1-\frac{\theta_t}{\sqrt{\theta^2_t+4p_t(1-p_t)\psi_t^{(\aleph)^2}}}\right)>\theta_t
\end{align*}
The PDE (\ref{eq_partial_49}) has Feynman-Kac probabilistic solution\footnote{Appendix E, formula (E.8) in Duffie (2001).}:
\begin{equation}
f(x,t) = \mathbb{E}^{\mathbb{Q}}[e^{-\int_t^T r_s ds}X_T|X_t = x],
\label{eq_fx_50}
\end{equation}
where the process $X_t,t\in[0,T]$ is defined on a stochastic basis $(\Omega,\mathbb{F} = \{\mathcal{F}_t\}_{t\geq 0},\mathbb{Q})$ with filtration $\mathbb{F}$, generated by the BM $B^{\mathbb{Q}}_t,t \geq 0$, and satisfies the stochastic differential equation:
\begin{equation}
d X_t = (r_t - D_{y,t}^{(\aleph,p;\mathcal{C})})X_t dt +\sigma_t X_t dB^{\mathbb{Q}}_t,t\in[0,T].
\label{eq_divi_51}
\end{equation}
The yield $D_{y,t}^{(\aleph,p;\mathcal{C})}$ for a misinformed trader is negative, and thus, in general, if  we set $p_{k \Delta t}^{\aleph} = \frac{1}{2}(1+\psi_{k \Delta t}^{(\aleph)}\sqrt{\Delta t})$, for some function $\psi_t^{(\aleph)}\in R, t\in [0,T]$, with continuous first derivative on $[0,T]$, the dividend yield $D_{y,t}^{(\aleph,p;\mathcal{C})}$  in (\ref{eq_divi_51}) is given by 
\begin{align*}
D_{y,t}^{(\aleph,p;\mathcal{C})} = \textup{sign} (\psi_t^{(\aleph)})\left(1-\frac{\theta_t}{\sqrt{\theta^2_t+4p_t(1-p_t)\psi_t^{(\aleph)^2}}}\right).
\end{align*}

\section{Conclusion}
In the literature on binomial option pricing, valuation is performed in four steps. In the first step,  the binomial model in the natural world is used, where the probability for the underlying stock upturn and stock mean return are model parameters.  Then, the risk-neutral probabilities are found, which should depend  on those two parameters, as shown in  Kim et al. (2016, 2019). The second step involve obtaining continuous-time model using the Donsker-Prokhorov invariance principle to derive the continuous-time dynamics of the underlying stock in the natural world. It is in this step that the probability for a stock upturn is naturally lost. Deriving the continuous-time risk-neutral valuation is the third step, where due to the presumed ability of the hedger to trade continuously with no transaction costs, the second parameter, the stock mean return, also disappears. In the fourth step, returning to the risk-neutral option price dynamics in the binomial discrete-time model, the risk-neutral probability now depends neither on the probability for a stock upturn nor on the mean return. In trinomial and multinomial option pricing models, the first three steps are abandoned, and only the last step is considered, leaving silent the issue of which discrete-pricing model in the natural world led to the discrete model in the risk-neutral world.

This four-step approach just described has obviously a very serious gap. That is, in the real world, no option trader who trades in discrete-time instances will disregard the information about the stock upturn probability and the stock mean return. In this paper, we derived option pricing models for traders with information on these two important parameters and we provided numerical illustrations which include the implied mean return surface and the implied surface for the probability for a stock upturn.  We derived our results when the pricing tree approximates a geometric Brownian motion, and more generally, a continuous-diffusion.
\section*{Reference}
\noindent E. Al\`{o}s, M. Mancino, \& T. Wang (2019) Volatility and volatility-linked derivatives: estimation, modeling, and pricing,
{\it Decisions in Economics and Finance} {\bf 42}, 321--349.\\[3pt]
\noindent A. Ang \& G. Bekaert (2007) Stock return predictability: Is it there? 
{\it Review of Financial Studies} {\bf 20}, 651--707.\\[3pt]
\noindent M. Andersson (1998) On Testing and Forecasting in Fractionally Integrated Time Series Models. Ph.D. thesis, 
{\it Stockholm School of Economics, Stockholm}.\\[3pt]
\noindent K. Back (1992) Insider trading in continuous time, 
{\it Review of Financial Studies} {\bf 5}, 387--409.\\[3pt]
\noindent K. Back (1993) Asymmetric information and options, 
{\it Review of Financial Studies} {\bf 6}, 435--472.\\[3pt]
\noindent K. Back \& Baruch S. (2004) Information in securities markets: Kyle meets Glosten and Milgrom, 
{\it Econometrica} {\bf 72}, 433--465.\\[3pt]
\noindent K. Back \& H. Pedersen (1998) Long-lived information and intraday patterns, 
{\it Journal of Financial Markets} {\bf 1}, 385--402.\\[3pt]
\noindent C. Boucher (2006) Stock prices–inflation puzzle and the predictability of stock market returns, 
{\it Economics Letters} {\bf 90}, 205--212.\\[3pt]
\noindent P. Billingsley (1965) {\it Ergodic Theory and Information}, Second Edition, John Wiley \& Sons, New York.\\[3pt]
\noindent P. Billingsley (1999) {\it Convergence of Probability Measures}, 
Second Edition, Wiley-Interscience, New York.\\[3pt] 
\noindent F. Black \& M. Scholes (1973) The pricing of options and corporate liabilities, 
{\it Journal of Political Economy} {\bf 81}, 637--654.\\[3pt]
\noindent M.K. Brunnermeier (2001) {\it Asset Pricing under Asymmetric Information: Bubbles, crashes, Technical Analysis and Herding},
Oxford University Press, Oxford.\\[3pt]
\noindent R. Caldentey \& E. Stacchetti (2010) Insider trading with a random deadline, 
{\it Econometrica} {\bf 78}, 245--283. \\[3pt]
\noindent J.Y. Campbell (2000) Asset pricing at the millennium. 
{\it Journal of Finance} {\bf 55}, 1515--1567. \\[3pt]
\noindent J.Y. Campbell, A.W. Lo \& A.C. MacKinlay (1997) {\it The Econometrics of Financial Markets},
Princeton University Press, Princeton. \\[3pt]
\noindent J.Y. Campbell \& M. Yogo (2005) Efficient tests of stock return predictability, 
{\it Journal of Financial Economics} {\bf 8}, 27--60.\\[3pt]
\noindent M.G. Caporale \& L.A. Gil-Alana (2014) Fractional integration and cointegration in US time series data, 
{\it Empirical Economics} {\bf 47}, 1389--1410. \\[3pt]
\noindent M. Caporin, A. Ranaldo \& P. Santucci de Magistris (2013) On the predictability of stock prices: A case for high and low prices, 
{\it Journal of Banking and Finance} {\bf 37}, 5132--5146.\\[3pt]
\noindent \c{S}. \c{C}elik (2012) Theoretical and empirical review of asset pricing models: A structural synthesis international, 
{\it Journal of Economics and Financial Issues} {\bf 2}(2), 141--178.\\[3pt]
\noindent R. Cervell\'{o}-Royo, F. Guijarro \& K. Michniuk (2015) Stock market trading rule based on pattern recognition and technical analysis: Forecasting the DJIA index with intraday data,
{\it Expert Systems with Applications} {\bf 42}, 5963--5975.\\[3pt]
\noindent P. Cheridito (2003) Arbitrage in fractional Brownian motion models, {\it Finance and  Stochastics} {\bf 7}, 533--553.\\[3pt]
\noindent K-H. Cho (2003) Continuous auctions and insider trading: Uniqueness and risk aversion, 
{\it Finance and Stochastics} {\bf 7}, 47--71. \\[3pt]
\noindent J. Cochrane (2001) {\it Asset Pricing}, 
Princeton: University Press. \\[3pt]
\noindent P. Collin-Dufresne \& V. Fos (2015) Do prices reveal the presence of informed trading? 
{\it Journal of Finance} {\bf 70}, 1555--1582. \\[3pt]
\noindent F. Comte \& E. Renault (1998) Long-memory in continuous-time stochastic volatility models, 
{\it Mathematical Finance} {\bf 8}, 291--323. \\[3pt] 
\noindent R. Cont (2001) Empirical properties of asset returns: Stylized facts and statistical issues, 
{\it Quantitative Finance} {\bf 1}, 223--236. \\[3pt]
\noindent R. Cont (2005) Long range dependence in financial markets, 
{\it Lévy-Véhel J. and Lutton E. (edt) Fractals in Engineering: New Trends and Applications}, 159--179. \\[3pt]
\noindent R. Cont, \& P. Tankov (2004) {\it Financial Modelling with Jump Processes}, 
Boca Raton: FL: Chapman \& Hall / CRC. \\[3pt]
\noindent T. Cover \& J. Thomas (2006) {\it Elements of Information Theory}, 
Second Edition, Wiley-Interscience, John Wiley \& Sons, Hoboken, New Jersey. \\[3pt]
\noindent J. Cox, S. Ross \& M. Rubinstein (1979) Options pricing:  A simplified approach, 
{\it Journal of Financial Economics} {\bf 7}, 229--263. \\[3pt]
\noindent K. Daniel \& S. Titman (1999) Market efficiency in an irrational world, 
{\it Financial Analysis Journal} {\bf 55}, 28--40. \\[3pt]
\noindent Y. Davydov \& V. Rotar (2008) On a non-classical invariance principle. 
{\it Statistics $\&$ Probability Letters} {\bf 78}, 2031--2038. \\[3pt]
\noindent F.X. Diebold \& A. Inoue (2001) Long memory and regime switching. 
{\it Journal of Econometrics} {\bf 105}, 131--159. \\[3pt]
\noindent M.D. Donsker (1951) An invariant principle for certain probability limit theorems, 
{\it Memoirs of the American Mathematical Society} {\bf 6}, 1--10. \\[3pt]
\noindent I.I Gikhman \& A.V. Skorokhod (1969) {\it Introduction to the Theory of Random Processes}, W.B. Saunders Company, Philadelphia. \\[3pt]
\noindent D. Duffie (2001) {\it Dynamic Asset Pricing Theory}, 
Third Edition, Princeton University Press, Princeton, New Jersey. \\[3pt]
\noindent E. Fama (1965) The behavior of stock market prices, 
{\it Journal of Business} {\bf 38} (b), 34--105. \\[3pt]
\noindent E. Fama (1970) Efficient capital markets: A review of theory and empirical work,  
{\it Journal of Finance} {\bf 25}, 383--417. \\[3pt]
\noindent U. Horst \& F. Naujokat (2011) On derivatives with illiquid underlying and market manipulation, 
{\it Quantitative Finance} {\bf 11}, 1051--1066. \\[3pt]
\noindent D. Hsieh (1990) Chaos and non-linear dynamics: Application to financial markets, 
{\it Journal of Finance} {\bf 46}, 1839--1878. \\[3pt] 
\noindent J. Hull (2012) {\it Options, Futures, and Other Derivatives}, Eighth Edition, Pearson, Boston. \\[3pt]
\noindent R. Jarrow \& A. Rudd (1983) {\it Option Pricing}, Irwin, Homewood, IL, Dow Jones-Irwin Publishing.\\[3pt]
\noindent F. Jovanovic \& C. Schinckus (2013) Econophysics: A new challenge for financial economics, 
{\it Journal of the History of Economic Thought} {\bf 35} (3), 319--352. \\[3pt]
\noindent B. Kelly \& A. Ljungqvist (2012) Testing asymmetric-information asset pricing models, 
{\it Review of Financial Studies} {\bf 25}, 1366--1413. \\[3pt]
\noindent Y.S. Kim, S.V. Stoyanov, S.T. Rachev \& F.J. Fabozzi (2016) Multi-purpose binomial model: Fitting all moments to the underlying Brownian motion, 
{\it Economics Letters} {\bf 145}, 225--229. \\[3pt]
\noindent Y.S. Kim, S.V. Stoyanov, S.T. Rachev \& F.J. Fabozzi (2019) Enhancing binomial and trinomial option pricing models, 
{\it Finance Research Letters} {\bf 28}, 185--190. \\[3pt]
\noindent A.S. Kyle (1985) Continuous auctions and insider trading. 
{\it Econometrica} {\bf 53}, 1315--1335. \\[3pt]
\noindent A.W. Lo (1991) Long memory in stock market prices. 
{\it Econometrica} {\bf 59}, 1279--1313. \\[3pt] 
\noindent A.W. Lo \& A.C. MacKinley (1988) Stock market prices do not follow random walks: Evidence from a simple specification test, 
{\it Review Financial Studies} {\bf 1}, 41--66. \\[3pt]
\noindent R. Merton (1973) Theory of rational option pricing, 
{\it Bell Journal of Economics and Management Science} {\bf 4}, 141--183. \\[3pt]
\noindent C. Munk (2013) {\it Financial Asset Pricing Theory}, Third Edition. New York: Oxford University Press. \\[3pt]
\noindent M. Nielsen (2010) Nonparametric cointegration analysis of fractional systems with unknown integration orders, 
{\it Journal of Econometrics} {\bf 155}, 1701--1787. \\[3pt]
\noindent M. Osborne (1959) Brownian motion in stock market, 
{\it Operation Research} {\bf 7}, 145--173. \\[3pt]
\noindent Yu. Prokhorov (1956) Convergence of random processes and limit theorems in probability theory, 
{\it Theory of Probability $\&$ Its Applications} {\bf 1}, 157--214. \\[3pt]
\noindent S. Rachev \& S. Mittnik (2000) {\it Stable Paretian Models in Finance}, Hoboken, NJ: John Wiley \& Sons. \\[3pt] 
\noindent S.T. Rachev, Y.S. Kim, M.L. Bianchi \& F.J. Fabozzi (2011) 
{\it Financial Models with L\'{e}vy Processes and Volatility Clustering}, Hoboken, NJ: John Wiley \& Sons. \\[3pt]
\noindent O. Rioul (2018) This is IT: A primer on Shannon’s entropy and Information, 
{\it L’Inforrmation, S\'{e}minaire Poincar\'{e}} {\bf XXIII}, 43--77. \\[3pt]
\noindent D. Robinson (2008) Entropy and uncertainty, 
{\it Entropy} {\bf 10}, 493--506. \\[3pt]
\noindent S. Rostek (2009) Option Pricing in Fractional Brownian Markets, 
{\it Lecture Notes in Economics and mathematical Systems} 622, Springer. \\[3pt]
\noindent M. Rubinstein (2001) Rational markets: yes or no? The Affirmative Case,
{\it Financial Analysts Journal} {\bf 57} (3), 15--29. \\[3pt]
\noindent W. Schoutens (2003) {\it L\'{e}vy Processes in Finance: Pricing Financial Derivatives}, New York: Wiley. \\[3pt]
\noindent R. Shiller (2003) From efficient markets theory to behavioral finance, 
{\it Journal of Economic Perspectives} {\bf 17} (1), 83--104. \\[3pt]
\noindent A. Shirvani, S.T. Rachev \& F.J. Fabozzi (2019) A Rational Finance Explanation of the Stock Predictability Puzzle. 
 arXiv:1911.02194\\[3pt]
\noindent A. Sidney (1961) Price movements in speculative markets: Trends of random walks, 
{\it Industrial Management Review} {\bf 2}, 7--26. \\[3pt]
\noindent A. Sidney (1964) Price movements in speculative markets: Trends of random walks, No2., 
{\it Industrial Management Review} {\bf 5}, 25--46. \\[3pt]
\noindent C. Skiadas (2009) {\it Asset Pricing Theory}, Princeton, NJ: Princeton University Press. \\[3pt]
\noindent A.V. Skorokhod (2005) {\it Basic Principles and Applications of Probability Theory}, Springer, Berlin.
\end{document}